%% file: main.tex
\documentclass{nature}
\usepackage{graphicx}
\usepackage{amsmath}
\usepackage{amssymb}
\usepackage{color}
\usepackage{url}
\usepackage{verbatim}
\usepackage{tabu}
\usepackage{sansmath}
\usepackage{upgreek}
\usepackage{epsfig}
\usepackage{setspace}
\usepackage{caption}
\usepackage{url}
\usepackage[dvipsnames]{xcolor}
\usepackage{amsmath}
\usepackage{amssymb}
\usepackage{sansmath}
\usepackage{gensymb}
\usepackage{hyperref}
\usepackage[margin=2cm,footskip=0.25in]{geometry}

\usepackage[symbol]{footmisc}
\makeatletter
\let\blx@rerun@biber\relax
\makeatother

\DeclareCaptionLabelFormat{adja-page}{\hrulefill\\#1 #2 \emph{(previous page)}}

\def\lapp{\ifmmode\stackrel{<}{_{\sim}}\else$\stackrel{<}{_{\sim}}$\fi}
\def\gapp{\ifmmode\stackrel{>}{_{\sim}}\else$\stackrel{>}{_{\sim}}$\fi}

\defbibheading{subbibliography}{%
  \section*{}}
\let\cite\autocite

\begin{document}

\begin{refsegment}
\defbibfilter{notother}{not segment=\therefsegment}

\begin{center}
    {\Large \bf On the detection of a cosmic dawn signal in the radio background}
\end{center}

\vspace{1cm}

\input{author_list}

\clearpage


\textbf{
The astrophysics of cosmic dawn, when star formation commenced in the first collapsed objects, is predicted to be revealed as spectral and spatial signatures in the cosmic radio background at long wavelengths\cite{2006PhR...433..181F}.  
The sky-averaged redshifted 21-cm absorption line of neutral hydrogen \cite{2012RPPh...75h6901P,2014Natur.506..197F} 
is a probe of cosmic dawn. 
The line profile is determined by the evolving thermal state of the gas, radiation background, Lyman-$\alpha$ radiation from stars scattering off cold primordial gas and the relative populations of the hyperfine spin levels in neutral hydrogen atoms \cite{1952AJ.....57R..31W,1959ApJ...129..536F,1959ApJ...129..551F}.
We report a radiometer measurement of the spectrum of the radio sky in the 55--85~MHz band, which shows that the profile found by Bowman et al. \cite{2018Natur.555...67B} in data taken with the Experiment to Detect the Global Epoch of Reionization Signature (EDGES) low-band instrument is not of astrophysical origin; their best-fitting profile is rejected with 95.3\% confidence.
The profile was interpreted to be a signature of cosmic dawn; however, its amplitude was substantially higher than that predicted by standard cosmological models \cite{2021arXiv210101777R}. Explanations for the amplitude of the profile included non-standard cosmology \cite{PhysRevLett.121.081305}, additional mechanisms for cooling the baryons, perhaps via interactions with millicharged dark matter \cite{2018Natur.555...71B} and an excess radio background \cite{Ewall_Wice_2018} at redshifts beyond 17.
Our non-detection bears out earlier concerns \cite{Hills2018,Singh_2019,Bradley_2019,10.1093/mnras/stz3388} and suggests that the profile found by Bowman et al.\cite{2018Natur.555...67B} is not evidence for 
new astrophysics or non-standard cosmology.
}

\vspace{0.5cm}
{\bf Radio sky spectrum measured using SARAS~3 radiometer.}
The Shaped Antenna measurement of the background RAdio Spectrum 3 (SARAS~3) experiment is a spectral radiometer \cite{2021ExANambissan,2020JAI.....950006G} based on a monocone antenna \cite{2021ITAP.Raghunathan} floated on a large body of water. SARAS~3 was deployed on lakes in Southern India during January--March 2020 and observations were made of the cosmic radio spectrum over the frequency band 43.75--87.5~MHz. The sites were selected by considering the characteristics of the lake water, depth, radio frequency environment including out of band FM, in relation to the system performance limitations.  Observations were carried out at night; the total electron content 
was found to be below $1.3 \times 10^{17}$~electrons~m$^{-2}$ during the observations \cite{CDDIS_Ionosphere} implying quiet ionospheric conditions \cite{2017isra.bookT,2014MNRAS.437.1056V}.
The data were calibrated and processed by algorithms that rejected data corrupted by terrestrial interference, then corrected for antenna efficiency, receiver noise temperature and thermal emission from water, and an averaged spectrum of the radio sky was formed by integrating over all times.  The instrument, algorithms, and processing steps are described in Methods. 

The analysis provided spectrophotometric measurement of the radio sky, in brightness temperature, over a useful science band 55 to 85~MHz, with spectral resolution and channel spacing of 61~kHz. Frequencies beyond the selected band boundaries had excessive flagging due to radio frequency interference. The time-averaged spectrum is shown in Fig.~\ref{fig:skyspectrum}; hereinafter we refer to this as the SARAS~3 spectrum. The radio spectrum is, as expected, close to a power law \cite{2017ApJ...840...33S} and has a best-fit temperature spectral index of $-2.42$.  We model the Galactic and extragalactic foregrounds in the band, propagated through the ionosphere, plus any systematic errors in calibration for receiver noise temperature, water thermal emission and antenna total efficiency, as a $6^{\rm th}$ order polynomial in log temperature versus log frequency. A discussion on systematic errors in the calibrations leading to the rationale for adopting this model is in Methods.
Also shown in Fig.~\ref{fig:skyspectrum} are residuals after subtracting out a best-fit $6^{\rm th}$ order polynomial, the distribution of the root mean square (RMS) value of measurement noise over frequency channels, and a plot of the residual spectrum normalised by these RMS values.  The calibrated sky spectrum, at native resolution of 61~kHz, has measurement noise of RMS value 213~mK. 

{\bf On the profile found by Bowman et al. \cite{2018Natur.555...67B} at 78~MHz.}
The Experiment to Detect the Global Epoch of Reionization Signature (EDGES)~\cite{2018Natur.555...67B} collaboration reported finding a feature in their spectrum that was modelled as a U-shaped profile, centred at $78 \pm 1$~MHz, with amplitude $500_{-200}^{+500}$~mK, full width at half amplitude of $19_{-2}^{+4}$~MHz and flattening parameter $\tau=7_{-3}^{+5}$.  Fig.~\ref{fig:skyspectrum} also shows their profile, along with frequency span of the SARAS~3 data and analysis presented here.
 
The SARAS~3 spectrum was first modelled with a $6^{\rm th}$ order polynomial to represent the foreground plus any calibration systematics. In Fig.~\ref{fig:1} we show the residuals from this model fit, which assumes the null hypothesis $H_{0}$ that the SARAS~3 spectrum does not have the feature found by Bowman et al. \cite{2018Natur.555...67B}.  The SARAS~3 residuals, smoothed to 1.4, 2.8 and 4.2~MHz resolutions, are also displayed and the RMS values of the residuals listed. It may be noted that the smoothed spectra retain the native channel spacing and spectral sampling of 61~kHz and are, therefore, oversampled. 
To provide expectations for RMS values of measurement noise in the smoothed SARAS~3 spectra, mock spectra were synthesised 
using the Global Sky Model (GSM) \cite{2008MNRAS.388..247D,10.1093/mnras/stw2525},
the SARAS~3 beam and with added Gaussian random noise with the distribution shown in panel (c) of Fig.~\ref{fig:skyspectrum}.  The mock spectra were processed to fit out $6^{\rm th}$ order polynomials and smoothed.  The expectations for RMS values of measurement noise for smoothed spectra, derived from the mock data, are listed in Fig.~\ref{fig:1} and show that the SARAS~3 measurement has no significant excess variance or any detection of spectral distortions in the radio sky, within the measurement uncertainty. 

Next, the SARAS~3 spectrum was jointly modelled  with a $6^{\rm th}$ order polynomial along with the best-fitting profile found by Bowman et al.\cite{2018Natur.555...67B}; the residuals from this modelling are also shown in Fig.~\ref{fig:1}. The assumption of hypothesis $H_{1}$ that the profile found by Bowman et al.\cite{2018Natur.555...67B} is a feature in the radio sky increases the residual variance, suggesting that the measurement using the SARAS~3 instrument prefers that the profile not be included in the sky model. Also shown in Fig.~\ref{fig:1} are corresponding residuals obtained when we synthesise mock spectra without measurement noise, using GSM for the foreground sky, and add the profile found by Bowman et al.\cite{2018Natur.555...67B}.  Modelling using a $6^{\rm th}$ order polynomial yields residuals that
show the signature expected, at resolutions of 1.4, 2.8 and 4.2~MHz, if the profile is present in the sky spectrum.  It may be noted here that although the detected profile has an amplitude of 500~mK, once our adopted foreground model has been fit to the sky spectrum and subtracted out the residual `processed signature' shown in Fig.~\ref{fig:1}c has an RMS value of only 19.4~mK. Naturally, for the adopted foreground model, the significance an inference of whether or not the profile found by Bowman et al.\cite{2018Natur.555...67B} is present in the SARAS~3 data is determined by this RMS value of the residual processed signature. 

The increase in the variance of the SARAS~3 residuals, when the sky model includes the profile, appears to be somewhat less than what would be expected if the data were uncorrelated with the profile; however, quantitative analysis described below shows that the deficit is consistent with statistical errors from measurement noise.

A quantitative measure of the presence of the profile found by Bowman et al.\cite{2018Natur.555...67B} in the SARAS~3 spectrum is derived using Markov Chain Monte Carlo (MCMC) analysis \cite{2013PASP..125..306F}. A joint fit was done of the SARAS~3 spectrum $T(\nu)$ with a model representing the foreground plus calibration systematics along with the profile.  The modelling is described by:
\begin{equation}
{\rm log_{10}} \{ (T(\nu)/{1~\rm K}) - s \times (T_{\rm EDGES}(\nu)/{1~\rm K}) \} = \sum\limits_{i=0}^{i=6} a_i \, \Re({\rm log_{10}}(\nu/{1~\rm MHz}))^{i},
\label{eq:model}
\end{equation}
\noindent where $\nu$ is frequency, $a_i$ are the coefficients of the $6^{th}$ order polynomial that models the foreground plus calibration systematics, $T_{\rm EDGES}(\nu)$ is the best-fitting profile that was found by Bowman et al.\cite{2018Natur.555...67B} and $s$ is a multiplying scale factor for the profile. The $\Re$ operator linearly re-scales values to be between $-1$ and $+1$. The $a_i$ coefficients and scale factor $s$ are free parameters. If the SARAS~3 spectrum of the sky contains the profile $T_{\rm EDGES}$, then we expect the MCMC analysis to yield a scale factor of unity. A scale factor of zero would imply that the profile is absent in the SARAS~3 spectrum. This approach has been used previously to examine for models for cosmic dawn and reionisation using data from the EDGES \cite{2017ApJ...847...64M} and SARAS \cite{0004-637X-858-1-54} instruments.  

Fig.~\ref{fig:MCMC_triangle} shows results of the MCMC analysis.  Convergence of the solutions was verified by examining the autocorrelation time and mean acceptance fraction. Fig.~\ref{fig:1D_dist} shows, separately, the 1D distribution of scale factor $s$ marginalized over other model parameters. MCMC analysis gives the most likely value of the scale factor $s$ to be $-0.01_{-0.51}^{+0.50}$, where the bounds are 1-standard deviation ($\sigma$) confidence intervals; this implies that the best-fitting profile found by Bowman et al.\cite{2018Natur.555...67B} is likely not there in the SARAS~3 spectrum. To these statistical errors in the scale factor, we add systematic errors corresponding to the uncertainties in the calibration corrections applied to the data, along with an additional error term representing 95\% confidence upper limits on any residual systematics after fitting out our foreground model (see Methods for details).  Accounting for systematic errors, the confidence in the best-fit scale factor $s$ is relaxed to be $-0.01_{-0.6}^{+0.6}$. The corresponding probability of false negative is 4.7\%, giving the significance of rejection of the hypothesis that the profile is present in the SARAS~3 measurement to be 95.3\%. 
We have also examined for the range of profiles allowed by the analysis \cite{2018Natur.555...67B} of the data from the EDGES instrument to explore the complete range of profile parameters admitted by the EDGES data. The MCMC analysis for the full range is discussed in Methods where the profile set as a whole is rejected with 90.4\% confidence. 

{\bf On the sky spectrum from the EDGES Low-band instrument.}
When the analysis of Bowman et al. \cite{2018Natur.555...71B}  subtracted out their best-fit foreground model from their measurement of the sky spectrum with the EDGES Low-band instrument, the residual spectrum had an RMS value of 87~mK \cite{2018Natur.555...71B} in which the measurement noise component was about 17~mK \cite{Singh_2019}. Thus, their measurement did have a spectral distortion with high significance.

We restrict to the common band of 55--85~MHz and fit a $6^{\rm th}$ order polynomial to the sky spectrum from the EDGES Low-band instrument \cite{EDGES_datarelease}
and also to that from SARAS~3, and compute the residuals.  These were smoothed to a lower and common resolution of 2.8~MHz to improve sensitivity for detection of any wide-band spectral distortions. The smoothed residuals are shown in Fig.~\ref{fig:comparison}.  Also plotted are residuals of the EDGES Low-band measurement at their native resolution, where we find that the RMS value of the residuals is 29~mK, again significantly greater than the measurement noise of 17~mK \cite{Singh_2019}.
The RMS value of the residual spectrum corresponding to the SARAS~3 spectrum reduces with smoothing as expected for Gaussian random noise, as shown in panel (b) of Fig.~\ref{fig:comparison}.  However, the RMS value 
of the residuals corresponding to data from the EDGES instrument declines less with smoothing. 

The above analyses show that the residual spectrum from the EDGES instrument obviously has a distortion feature inconsistent with measurement noise.  To make a quantitative estimate of whether the spectral distortion in the measurement using the EDGES Low-band instrument is present in the SARAS~3 spectrum, we compute a normalised correlation $\zeta$.  This statistic corresponds to matched filtering of the residuals of the SARAS~3 spectrum using the corresponding residuals of the EDGES Low-Band measurement as template:
\begin{equation}
\zeta = \frac{{\mathbb E}[SE]}{{\mathbb E}[E^2]},
\end{equation}
\noindent where $E$ is the residual from the sky spectrum made using the EDGES Low-band data, $S$ is the residual from SARAS~3, and operator ${\mathbb E}$ denotes expectation that is obtained by averaging over frequency.   
We have simulated mock data, with the measurement noise present in the spectrum made using the SARAS~3 instrument, to derive the expected distributions of $\zeta$ for two cases: hypothesis $H_0$ in which the sky spectrum has no spectral distortions and $H_1$ in which the sky spectrum has a residual distortion corresponding to that in the spectrum made with the EDGES instrument. The simulations confirm that in the case of the null hypothesis $H_0$, $\zeta$ is expected to be zero, and in the case of $H_1$, $\zeta$ is expected to be unity.  

The value of $\zeta$ is 0.12 for the pair of residual spectra, at 2.8~MHz resolution, shown in panel (a) of Fig.~\ref{fig:comparison}.  The simulations yield the distributions expected for the two hypotheses and these are also shown in Fig.~\ref{fig:comparison}.  The distributions in $\zeta$ in the two cases have standard deviations of 0.43 and give the probability of a false negative to be 2\%. Since we are directly cross-correlating two different measurements, this analysis yields improved constraints compared to the joint fitting approach described in the previous section. 

In summary, the correlation analysis demonstrates that the distortions in the
spectrum made using the EDGES Low-band instrument is, with 98\% confidence, not present in the SARAS~3 spectrum. 

{\bf The origin of the profile found by Bowman et al.\cite{2018Natur.555...67B}}
Independent analysis\cite{Hills2018} of the spectrum measured by Bowman et al.\cite{2018Natur.555...67B} indicated that there might be an uncalibrated systematic in the form of a sinusoidal undulation with period 12.5~MHz. 
It has also been shown \cite{Singh_2019} that allowing for such a systematic is preferred by Bayesian information criterion \cite{1978AnSta...6..461S} and obviates the requirement that the cosmic dawn signal be beyond predictions of standard cosmological models.
Inhomogeneous ground beneath the antenna of the EDGES Low-band instrument has been suggested\cite{Bradley_2019} as an origin for systematic errors. 

The SARAS~3 instrument differs significantly from EDGES in that the antenna is a short monocone and that the sensor is on a large water body, thus considerably reducing beam chromaticity and inhomogeneities in environment as potential sources of systematic errors.  
The profile found by Bowman et al.\cite{2018Natur.555...67B} is not detected in the MCMC analysis of the sky spectrum made with the SARAS~3 instrument.  Moreover, the correlation analysis shows that the distortion present in the spectrum made using the EDGES Low-band instrument, which was used
to derive the best-fitting profile and define the bounds on the parameter space for the profile, is not present in the SARAS~3 spectrum of the sky.  These suggest that the significant spectral distortion present in the sky spectrum made with the EDGES Low-band instrument is a systematic associated with the instrument.   The  SARAS~3 spectrum and its analysis, which we present here, thus lends weight to earlier independent analyses\cite{Singh_2019}$^,$\cite{10.1093/mnras/stz3388} that argued that the sky spectrum made with the EDGES Low-band instrument requires modelling with a foreground model and also a formulation for systematics.

Linear polarisation in Galactic foreground, together with inevitable Faraday Rotation, would result in confusing spectral structure in radiometers like EDGES and SARAS~3, which have single linearly polarized antennas \cite{2019MNRAS.489.4007S}. These polarization systematics would differ in the two instruments.  
The extent to which measurements with the EDGES and SARAS~3 instruments are confused by polarisation systematics would be discerned with improved sensitivity and deploying at different sites, to examine for changes in any detected spectral structure with local sidereal time and site latitude.

The sensitivity of the SARAS~3 data rules out a cosmological origin for the profile found by Bowman et al.\cite{2018Natur.555...67B} and suggests that the spectral distortions in the measured sky spectrum by the EDGES low-band instrument is dominantly instrument systematics. Avoidance of systematics requires careful control of internal receiver systematics and deployment in environs that are electromagnetically homogeneous and without back scatter. We conclude that continued observations with sensors deployed in such environs, like the SARAS~3 monocone on large water bodies in remote locales on Earth or a space mission in orbit on the lunar farside \cite{2017ApJ...844...33B}, would provide data free from systematics and lead to discovery of the true redshifted 21-cm signal from cosmic dawn.

\vspace{1cm}

\vspace{0.5cm}

\noindent {\bf Acknowledgements.} Authors would like to thank the staff at the Gauribidanur Field Station for assistance with system tests and measurements, and the Mechanical and Electronics Engineering Groups at the Raman Research Institute for building, assembling and deploying SARAS~3. S.S. thanks Jordan Mirocha and Adrian Liu for useful discussions on global signal modelling and acknowledges support from McGill Astrophysics postdoctoral fellowship. J.N.T. acknowledges a PhD bursary from the Australian Research Council (ARC) Future Fellowship under grant FT180100196. 

\vspace{0.5cm}

\noindent {\bf Author contributions.} R.S. led the experiment and supervised all stages of this work. N.U.S contributed to all activities of the work and provided expert feedback. S.S. performed the data selection, modelling and signal inference. J.N. formulated analogue receiver calibration \& characterisation, and contributed to data analysis. B.S.G, K.S.S, R.Som. and A.R. engineered various subsystems of the experiment, and participated in system testing and deployment. M.S.R. provided physically motivated sky model and formulated maximally smooth functions, used for qualifying the system. 

\vspace{0.5cm}

\noindent {\bf Author information.} The authors declare no competing interests. Correspondence and requests for materials should be addressed to saurabhs@rri.res.in.

\printbibliography[heading=subbibliography,segment=\therefsegment]
\end{refsegment}

\clearpage

\input{figures}

\clearpage

\begin{refsegment}

\input{methods}

\printbibliography[heading=subbibliography,segment=\therefsegment,filter=notother]
\end{refsegment}  

\clearpage

\input{extended_data}

\end{document}

%% file: author_list.tex
{\large Authors: Saurabh~Singh$^{1,2,3*}$, 
Jishnu~Nambissan T.$^{1,4}$, 
Ravi~Subrahmanyan$^{1,5}$,\\ 
N.~Udaya~Shankar$^{1}$,
B.~S.~Girish$^{1}$, 
A.~Raghunathan$^{1}$,
R.~Somashekar$^{1}$,
K.~S.~Srivani$^{1}$ 
\& Mayuri~Sathyanarayana~Rao$^{1}$
} \\

\vspace{0.3cm}

\noindent $^{1}$ Raman Research Institute, C V Raman Avenue, Sadashivanagar, Bangalore 560080, India. \\
$^{2}$ Department of Physics, McGill University, 3600 rue University, Montr\'eal, QC H3A 2T8, Canada. \\
$^{3}$ McGill Space Institute, McGill University, 3550 rue University, Montr\'eal, QC H3A 2A7, Canada. \\
$^{4}$ International Centre for Radio Astronomy Research, Curtin University, Bentley, WA 6102, Australia \\
$^{5}$ Space \& Astronomy, CSIRO, PO Box 1130, Bentley, WA 6102, Australia \\
$\ast$ Corresponding author. Email: saurabhs@rri.res.in

%% file: figures.tex
\begin{figure}
    \centering
    \includegraphics[width=0.7\textwidth]{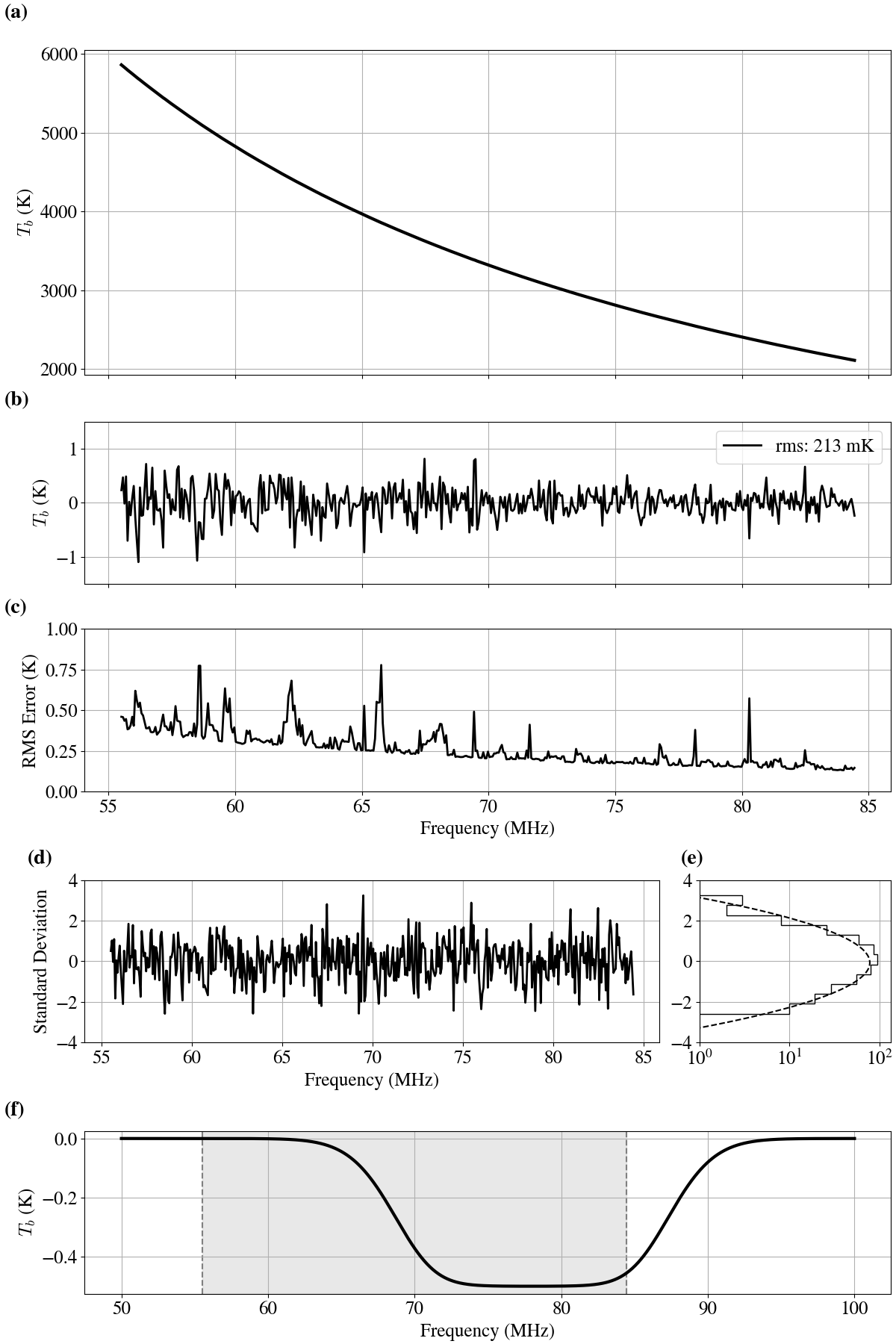}
    \caption{{\bf Spectrum of the radio sky.}
    The time-averaged spectrum of the radio sky as measured by the SARAS~3 radiometer is shown in panel (a).  Panel (b) shows residuals on subtracting out a best-fit $6^{\rm th}$-order polynomial model. Panel (c) shows the RMS value of measurement noise, at the native spectral resolution of 61~kHz, versus frequency.  Panel (d) shows the residuals with the value in each channel normalised by the RMS value of measurement noise in that channel, thus giving the residuals units of standard deviation. The histogram in panel (e) shows the distribution of normalised residuals in logarithmic scale; a best-fit parabola is overlaid. For reference, panel (f) shows the best-fit profile found by Bowman et al. \cite{2018Natur.555...67B}; the shaded region represents the frequency band of the SARAS~3 data and analysis described here. 
    }
    \label{fig:skyspectrum}
\end{figure}

\begin{figure}
\begin{center}
\includegraphics[width=1\textwidth]{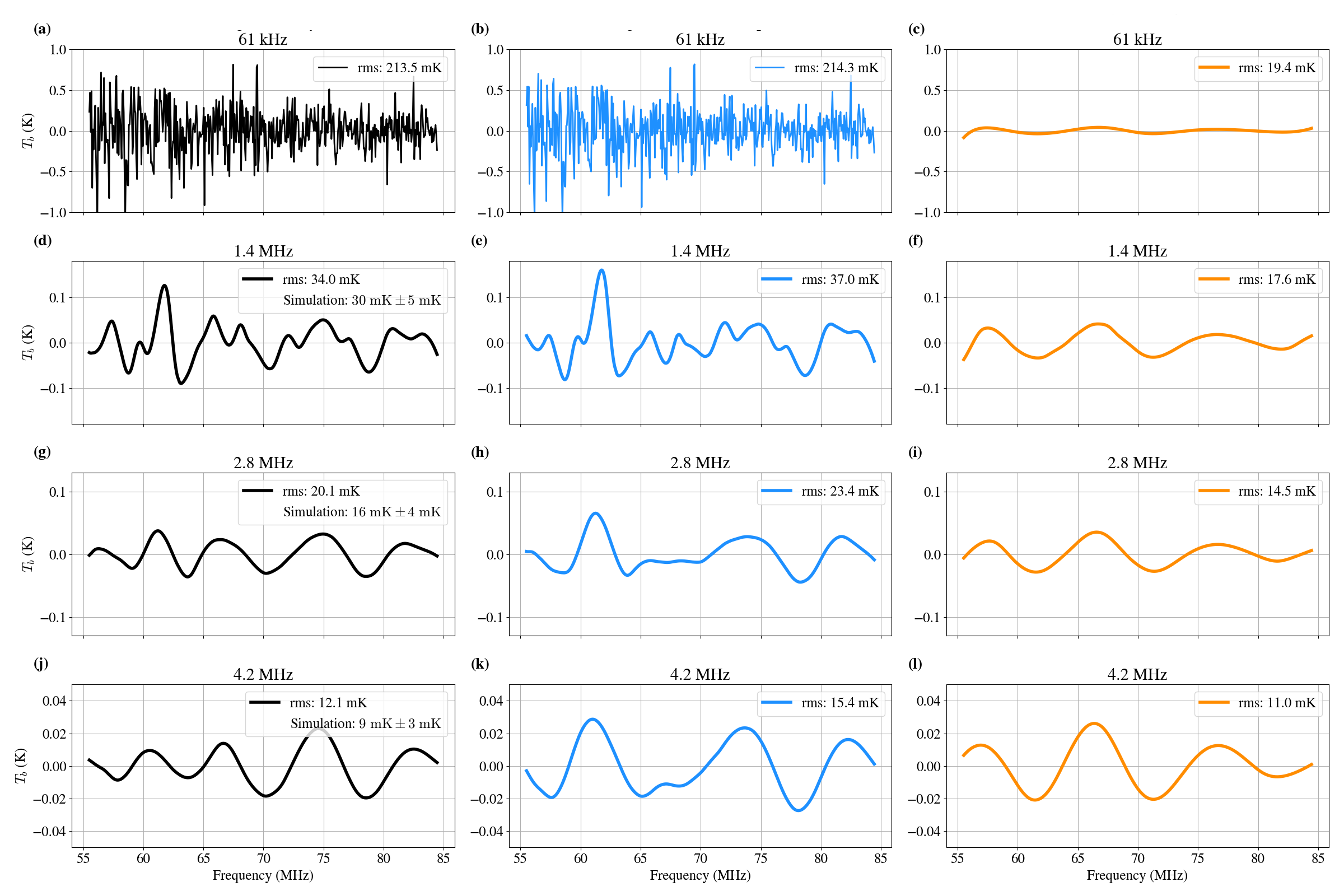}
\caption{{\bf Comparison of residuals.} 
The rows show residuals at native resolution of 61~kHz and smoothed to 1.4, 2.8 \& 4.2~MHz. Panels (a), (d), (g) and (j) in the left column show residuals when the measured spectrum is modelled with a $6^{\rm th}$ order polynomial. 
Panels (b), (e), (h) and (k) in the middle column show residuals when the 
model includes the best-fitting U-shaped profile found by Bowman et al.
 \cite{2018Natur.555...67B}. 
Panels (c), (f), (i) and (l) in the right column show residuals when mock sky spectra constructed using GSM sky plus the profile are modelled with a $6^{\rm th}$ order polynomial. The RMS values of the traces are in the legend in each panel; also included in the panels in the left column are RMS values of measurement noise, along with and $1\sigma$ confidence intervals, as derived from mock data.}
\label{fig:1}
\end{center}
\end{figure}

\begin{figure}
\centering
\includegraphics[width=1.0\linewidth]{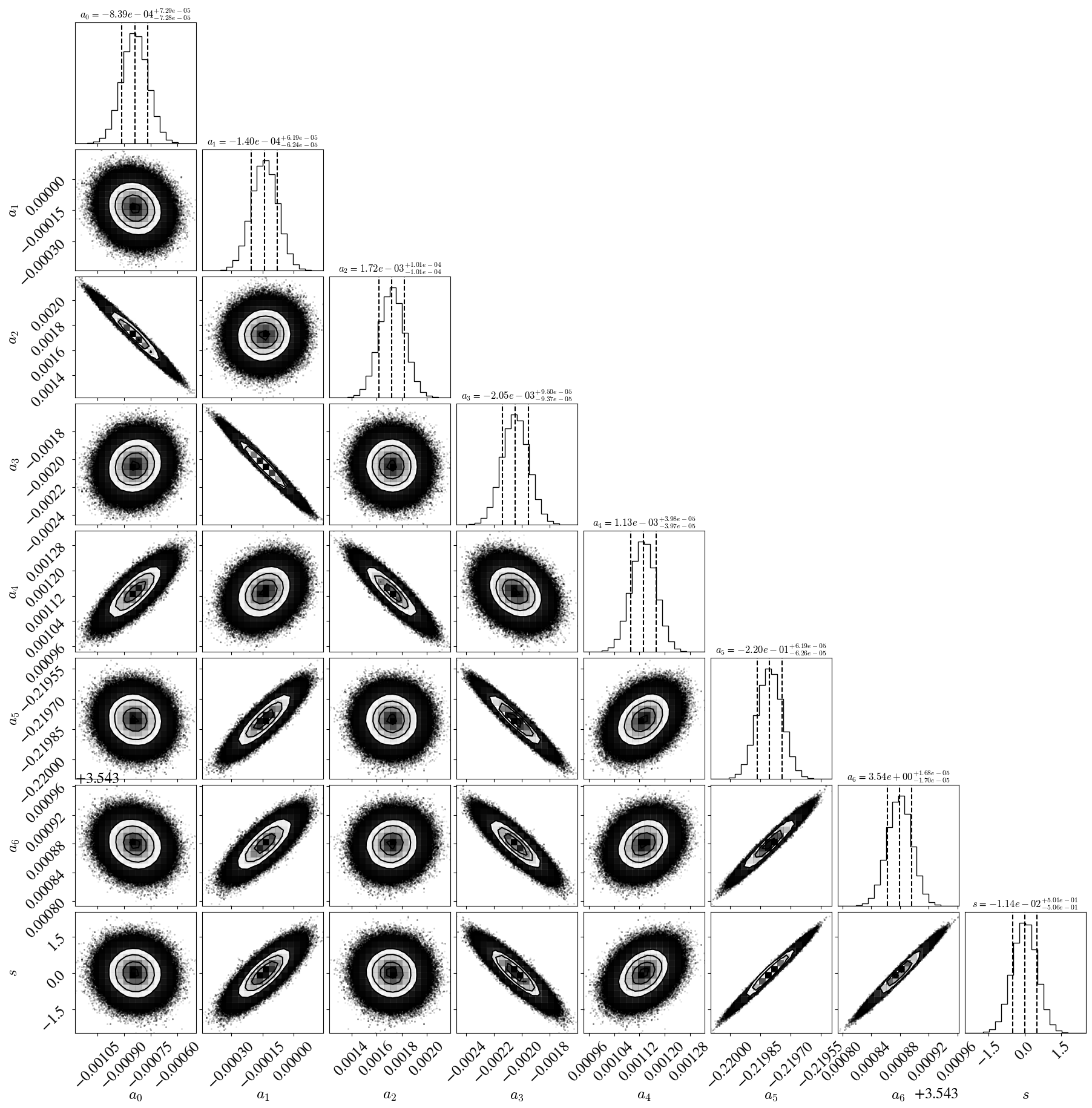}
\caption{{\bf MCMC modeling of the SARAS~3 spectrum.} 1D and 2D distribution of model parameters---coefficients $a_i$ representing the foreground polynomial and scale factor $s$---are shown. The axis range for each parameter has been set separately to span the range sampled by the MCMC walkers after the burn-in steps are discarded. The spread in the posteriors of the parameters represent statistical errors, driven by measurement noise.}
\label{fig:MCMC_triangle}
\end{figure}

\begin{figure}
\centering
\includegraphics[width=0.8\linewidth]{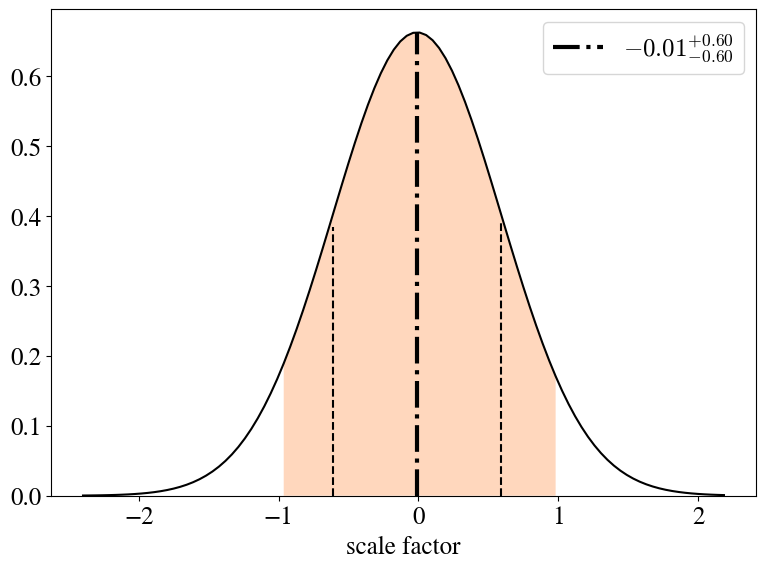}
\caption{{\bf 1D distribution of the scale factor $s$.}  The black dash-dotted line shows the mean value of the scale factor given by the MCMC analysis; its value and $1 \sigma$ error, including systematic errors, are in the legend. Dashed black lines represent $1\sigma$ confidence intervals. The shaded region spans $5^{\rm th}-95^{\rm th}$ percentile values. }
\label{fig:1D_dist}
\end{figure}

\begin{figure}
\centering
\includegraphics[width=0.4\linewidth]{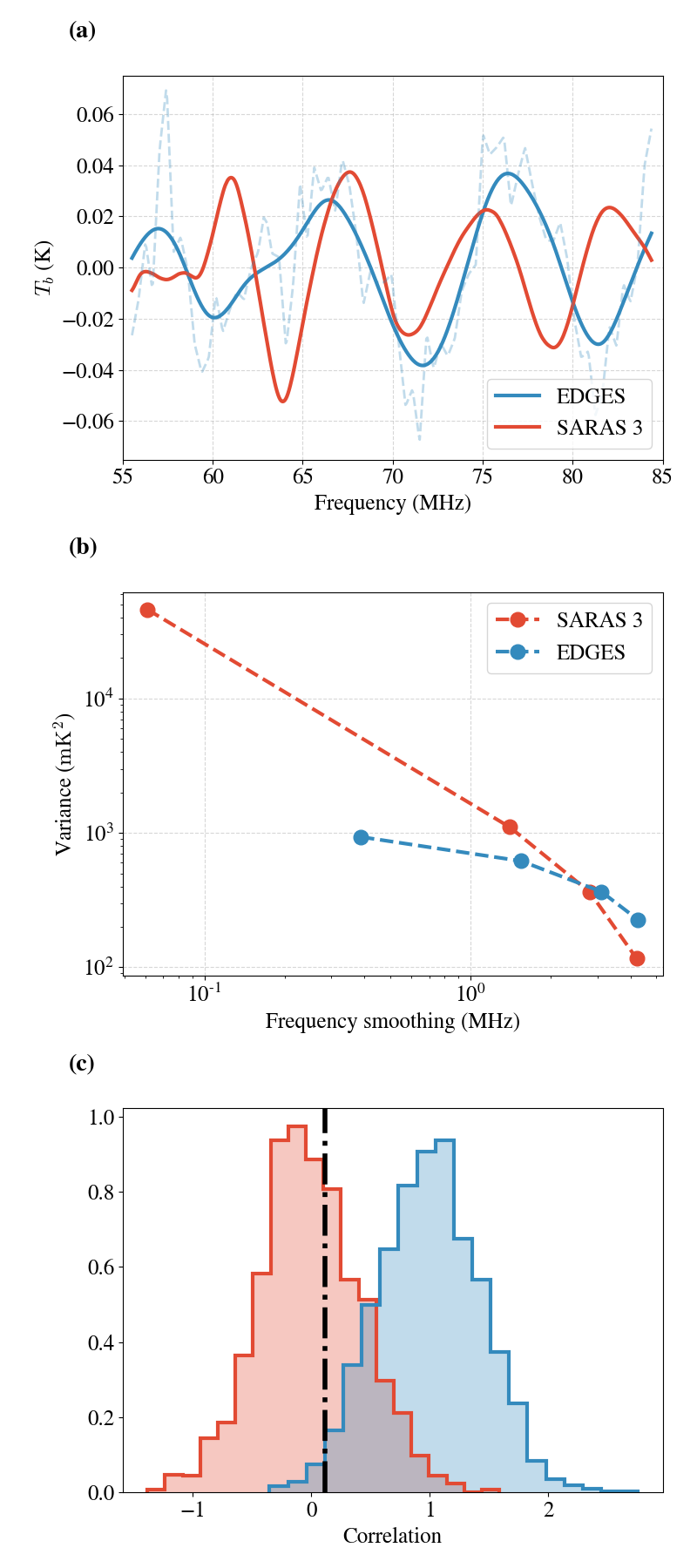}
\caption{{\bf Comparison of residuals from modelling of spectra from SARAS~3 and EDGES low-band instruments.} Panel (a) shows the residuals obtained on fitting out $6^{th}$-order polynomials from the EDGES and SARAS~3 measurements of the sky spectrum.  Residuals of spectra from both instruments are shown smoothed to a common resolution of 2.8~MHz, as continuous traces. Overlaid is the residual of the spectrum from the EDGES instrument, at their native resolution of 0.4~MHz, as a dashed blue trace. 
Panel (b) shows the variance in residuals versus resolution, on smoothing. 
Panel (c) shows distributions of correlation derived from mock data: in blue is shown the case where mock SARAS~3 data has a residual distortion corresponding to that in the spectrum made with the EDGES instrument, and in red is shown the case where mock SARAS~3 data have no such spectral distortion. The vertical dot-dashed line in this panel represents the correlation coefficient $\zeta$ derived from residuals of spectra from SARAS~3 and EDGES low-band instruments. }
\label{fig:comparison}
\end{figure}

%% file: methods.tex
\begin{center}
    {\Large \bf Methods}
\end{center}
\vspace{0.5cm}

\noindent {\bf The instrument.}  The SARAS~3 instrument is the third improved variant in the SARAS series of radiometers. These are spectral radiometers operating in a part of the 40-230~MHz band, consisting of an antenna, an analogue receiver and a digital spectrometer. The earliest version, SARAS~1, used a shaped fat-dipole over an absorptive ground plane~\cite{2013ExA....36..319P} and improved the calibration accuracy of previous 150-MHz all-sky map \cite{0004-637X-801-2-138,1970AuJPA..16....1L}. The second version used an electrically short spherical monopole \cite{Singh2018} and ruled out a class of Cosmic Dawn (CD)/Epoch of Reionization (EoR) models that featured weak X-ray heating and rapid reionization \cite{2041-8205-845-2-L12,0004-637X-858-1-54}.  Successive versions improved on the receiver architecture, calibration techniques, and adopted better methods for suppression of systematics. 

SARAS~3 observes over the 43.75--87.5 MHz band.  The electromagnetic sensor of the instrument is a monocone antenna \cite{2021ITAP.Raghunathan} floated on a large water body, as shown in Extended Data Figure~1. 
The slant height of the cone equals the radius of the metallic reflector plate; both these are 0.8~m thus making the radiating elements smaller than quarter wavelength at the highest operating frequency.   At selected observing sites, the water beneath the antenna provides a high permittivity medium that is homogeneous to beyond the penetration depth for radio waves in the observing band. Since the speed of propagation of electromagnetic (EM) waves in water in the operating band is about a factor 9 less than that in free space, a depth exceeding 12~m ensures that any spectral structures arising from reflections of sky or receiver noise temperature from the bottom would have periods less than 1.5~MHz, and thus not confuse CD/EoR signals that are predicted to be substantially broader in frequency.  In lakes where the water has moderate conductivity, there is significant attenuation for EM waves in the operating band and reduced depth suffices for the science goal.  The electrically small monocone over a medium that has high impedance contrast with free space results in close to frequency-independent beam, high efficiency, and extremely smooth reflection and radiation efficiencies; all these facilitate CD/EoR detection.   An added advantage of adopting a monocone design is that the antenna signal at the vertex of the cone is unbalanced, and simply continues as a coaxial transmission to the receiver without any balun.  The antenna is floated using a raft made of Styrofoam, which is transparent to EM waves in the operating band. Most of the construction of the conducting surface of the monocone is with aluminium sheets that are joint along lines across which currents are not expected, given the antenna symmetry; nevertheless, all joints are covered with conductive aluminium tape.  We have done EM modelling of the SARAS~3 antenna using WIPL-D and FEKO software, adopting the salinity of the water at the observing site. The beam is omnidirectional and close to achromatic, has nulls towards horizon and zenith, and peak response that has a small shift in elevation from about $26\degree$ to $24\degree$ between 45 and 85~MHz.  

The receiver architecture \cite{2021ExANambissan} is a hybrid between total-power and correlation spectrometers, with double differencing for improved cancellation of internal systematics; a schematic is shown in Extended Data Figure~2.  The Dicke receiver switches the first low-noise amplifier between the antenna and a reference load. Additionally, following initial amplification, the analogue signal is split, phase switched, and processed in parallel chains and then channelised in a digital correlation spectrometer \cite{2020JAI.....950006G}. The hybrid architecture provides band-pass calibration accuracy via the Dicke switch, high sensitivity by placing low-noise amplifiers early in the signal path as in total-power receivers, and also provides excellent suppression of internal systematics through the double differencing in the Dicke switch, phase switch and correlation receiver architecture. 

The SARAS~3 receiver is in part enclosed in an EM-shielded water and weather proof aluminium box attached below the antenna ground plane, while the other part is located at the base station, along with digital electronics.  
Radio Frequency over fibre (RFoF) is used for signal transport. 
The total electrical path length from the antenna terminals to the electrical to optical converter is designed so that multi-path reflections of system noise within the receiver does not produce spectral structure that might confuse any CD/EoR signals. 
With RF and control signals carried on fibre between the receiver system at antenna and base station on shore, there are no conducting cables connected to the antenna, thus providing electrical isolation of the antenna base electronics from rest of the signal chain and avoiding parasitic elements that might compromise the antenna performance.

The electronics at the base station 
is located on shore about 150~m from the antenna.  Optical detectors convert the radiometer signal back to RF domain and a cross-over switch connects the RF signal to a 180$\degree$ hybrid, which provides copies of the RF signal to a pair of receiver chains. The copies are in-phase or out of phase depending on the state of the cross-over switch.  
Elliptic filters provide sharp cutoff outside the observing band.
Offline differencing of the spectra recorded by the correlation spectrometer in the two states of the cross-over switch cancels unwanted and inevitable `board noise' present in the samplers due to high-speed clocks in the digital electronics. 
The SARAS~3 digital correlation spectrometer \cite{2020JAI.....950006G} is based on precision SPECtrometer (pSPEC) board, which digitises a pair of analogue signals, performs 16384-point windowed discrete Fourier transforms, and computes averaged complex cross-correlation spectra. 
The sampling in time is at 500~MHz.  8192-point spectra are recorded over 250~MHz with a spectral resolution of 61~kHz; the spectra are oversampled by a factor of 2 in frequency domain.
The pSPEC board, laptop, and digital circuits associated with timing and clock generation, together with associated switched-mode power supplies, are all in a custom designed enclosure with high EM shielding to prevent radiative mode radio frequency interference (RFI) leakage to the ASCU and antenna. Field measurements confirm that the EM shielding plus 150~m space loss between the base station and antenna together provide greater than 100~dB suppression and limits the magnitude of self-generated RFI picked up by the antenna to below 1~mK.  

Performance measurements of the digital spectrometer give its spurious-free dynamic range as 50~dB; this limitation is primarily due to the use of 10-bit samplers in the pSPEC board. Persistent and strong RFI at the observing sites where SARAS~3 was deployed is observed to be primarily from distant FM stations.  At the site, we measured the strength of FM received by the SARAS~3 antenna and receiver with and without the elliptic filters in the analogue chains: the strength of FM was 25~dB above sky brightness temperature without filters, and was cut down to 10~dB below sky brightness with the filters. The elliptic filters in the analogue chains together with the dynamic range of the digital spectrometer limit the amplitude of spurious features from FM to be below 1~mK in the observing band, and hence the site and system are jointly qualified as adequate for the science.

The combination of monocone design, floating on water, electrically small antenna, avoidance of a balun, placing of the receiver at the antenna terminals, optical isolation, extremely small electrical path between the antenna and optical modulators, all together provide close to achromatic beam, smooth emission contribution from water and smooth transfer functions for the sky spectrum and receiver noise.  Thus, to the accuracy needed for CD/EoR signal detection, many radiometer characteristics of SARAS~3 do not require accurate characterisation measurements in the laboratory or in the field. This is substantiated in discussions below and in Extended Data Figure~9 where substantial variations in the estimation of calibrations are shown to result in residuals with RMS values less than 1~mK.
This is a fundamental difference between the SARAS~3 design philosophy and that of the EDGES \cite{2018Natur.555...67B} instrument.

{\noindent \bf Calibration.} The reference load is a series attenuator connected to a wideband precision noise source; switching the noise source on and off, when the reference load is connected to the receiver, provides a means for calibration of the bandpass. 
The source itself is selected to be a precision GHz-bandwidth noise source, with high noise power and with internal compensation for immunity to variations in ambient temperature and control voltage.  Coupling to the receiver is via a high 23-dB attenuation that provides stability in matching so that the reference termination maintains its impedance match when the noise is on and off. The physical temperature of the reference load is logged with 0.15~K accuracy throughout the observation; these are available as metadata and used in the post-processing to derive absolute antenna temperatures from the Dicke-switched difference measurements.

The instrument is cycled through a set of states while observing: equal time is spent with receiver connected to (i) the antenna, (ii) reference load with calibration noise off and (iii) reference load with calibration noise on. In each of these, data are recorded separately with the correlation receiver phase switched.  Offline, data in the phase switch states are differenced first, then a second difference is computed between the antenna and reference state with noise source off. Separately, bandpass calibration data are derived as a difference between reference load with calibration noise source on and that with the noise source off.  Thus, in a cycle time of 8.23~sec, a set of six spectra corresponding to six states are recorded, and calibration involves correcting the double differenced data for bandpass, scaling to set the units to be in kelvin noise temperature, and adding the temperature of the reference, to yield a calibrated spectrum of the antenna temperature.

Absolute calibration of the receiver was determined in a laboratory measurement by replacing the antenna with a matched load, and placing precision temperature probes on this load and also the reference load.  
The load replacing the antenna was immersed in ice water and separately in a heated bath, allowed to stabilise and then slowly drift towards ambient, while data were recorded by the instrument.  Fits to the data, allowing for offset errors between the probes and noise temperatures, yielded accurate estimates for the effective temperature step of the calibration noise source. 
The absolute accuracy in the calibrated spectrum of the antenna temperature is set by (i) the absolute accuracy in the measurement of noise temperature of the reference termination, which is given by the offset term, 2.3~K, in the above fit, and (ii) the error in the scaling to kelvin that is given by the maximum possible error in the slope of the fit: $\leq$~1\%. The error in scaling dominates and hence the SARAS~3 measurements of antenna temperature are deemed to have absolute accuracy better than 1\%. 

{\noindent \bf Laboratory verification of the receiver system.} 
The SARAS~3 antenna is designed to present a smooth reflection coefficient for a receiver system that is of $50~\Omega$ impedance.  
However, the receiver input has a complex reflection coefficient that is different.  
Consequently, both the antenna temperature and also the receiver noise may suffer multiple internal reflections between the impedance mismatch at the antenna terminals and the impedance discontinuities in the receiver signal path, which may result in the measurement of the total system temperature acquiring confusing spectral structure.

In the laboratory, the antenna is replaced with a network made of discrete components that model the antenna reflection coefficient, so that the internal systematics in laboratory tests might be similar to that expected in field deployments with the antenna. The network model, henceforth called antenna simulator, includes an external precision $50~\Omega$ termination. The modular network is connected to the receiver and the system is enclosed in an aluminium box to shield from local RFI in the laboratory.   Apart from presenting to the receiver a reflection coefficient that resembles the antenna, the antenna simulator also provides additional noise from the $50~\Omega$ termination, which performs the function of the radiation resistance of an antenna. The contribution of this noise source to system temperature is shaped by the transfer function corresponding to the reflection coefficient of the antenna simulator. The resistive termination is at ambient temperature and provides a flat-spectrum noise; therefore, the simulator does not provide a noise equivalent to the power-law form sky spectrum that would be observed by the antenna when deployed. Nevertheless, this test provides a useful method to verify the receiver system performance.

With the antenna simulator replacing the antenna, data were recorded for 33~hr with the receiver system, digital correlation spectrometer and its 150-m fibre cable interconnects configured as for field observations, and the system switching through states.  The laboratory test measurement data were calibrated and reduced using the same software codes used for sky data. 
The calibrated spectrum was fit with a maximally smooth (MS) function \cite{2017AJ....153...26S,2021MNRAS.502.4405B}; the fit, measurement and residuals are shown in Extended Data Figure~3. 

The residuals were smoothed from the native resolution of 61~kHz using Hann functions \cite{1978IEEEP..66...51H} with full width at half maximum (FWHM) that increases by factor 2 till 3.9~MHz. Variance of the residuals decreases from 16.9~mK at the native resolution to 2.3~mK, consistent with expectations for instrument measurement noise, and no obvious structure is seen in the residuals when smoothed.  We infer that internal systematics in the SARAS~3 receiver, if present, have RMS value below about 0.9 mK, which is the nominal excess RMS amplitude of the smoothed residual.

{\noindent \bf Antenna reflection efficiency.} The SARAS~3 antenna is passive and hence reciprocal; additionally, the antenna structure is lossless in that it has no lossy balun and all radiating elements are made of aluminium. Therefore,  we have configured the antenna as a radiator and measured its complex voltage reflection coefficient $\Gamma$ and derived the reflection efficiency $(1-|\Gamma|^2)$.

The receiver below the antenna ground plane, within the raft, was replaced with a purpose-built system for measurement of reflection efficiency \cite{2021ITAP.Raghunathan}.  
The measurement includes effects of all environmental influences, including the lake bed.  
On a lake with depth of 7~m, the reflection efficiency shows a sinusoidal ripple with RMS value 0.0000125 and period 2.5~MHz, consistent with the depth. For the science data, the site---salinity and depth at the point of deployment---was selected to reduce the amplitude of any such a ripples to a level where the corresponding structure in measurement of radio sky would be less than 1~mK.

In Extended Data Figure~4 we show reflection efficiency measurement at a site selected for science data along with a fit using an MS function. The RMS value of the residual is $5 \times 10^{-6}$ at the spectral resolution of 0.7~MHz. The residual is consistent with measurement noise and qualifies the antenna in its environment for spectral measurements with this fractional accuracy.

{\noindent \bf Antenna radiation efficiency and total efficiency.} While the antenna voltage reflection coefficient $\Gamma$ and its reflection efficiency $(1-|\Gamma|^2)$ determine the spectral nature of receiver noise in the SARAS~3 measurement, the total efficiency of the antenna determines the antenna transfer function for the sky signal.  

In the SARAS~3 design, there is no extended metallic ground screen.  This design choice is motivated by the reasoning that for precision 21-cm cosmology at long radio wavelengths any ground screen would have to be extremely large and be of extremely low transmission so as not to admit confusing spectral structures from ground noise, edge reflections and chromatic beam. In SARAS~3, with the small metallic ground plane of 0.8-m radius, radiation efficiency has been compromised for spectral smoothness. Therefore, the total efficiency of the antenna differs significantly from its reflection efficiency and is a product of the radiation and reflection efficiencies. Measurements of sky spectra with the instrument are a product of the true sky spectra with total efficiency, along with thermal emission from water that is shaped by the characteristics of radiation efficiency, and also thermal emission from resistive losses in the signal chain leading to the Dicke switch in the receiver. 

To measure the total efficiency, we employ a technique \cite{Singh2018} similar to that used for SARAS~2.   
The measurement data used here are reduced data---the data reduction is described below---calibrated for receiver bandpass and cleansed of RFI.  The reduced data represents the spectral distribution of antenna temperature versus local sidereal time (LST) along with receiver noise and thermal emission from water, which are expected to be stable over time within that due to drifts in ambient temperature.  Global sky model (GSM) \cite{10.1093/mnras/stw2525} is convolved with the chromatic SARAS~3 beam pattern to predict sky spectra that would be measured by the instrument, at any LST, assuming unity total efficiency.  
The horizon limit of the sky takes into account blockage by surrounding terrain. At each frequency, the distribution of antenna temperature versus predicted sky spectral brightness is fit with a straight line. The fits weight the data points to take into account the varying averaging time due to data rejected as RFI.  The slope of the fit yields the total efficiency.  

The total efficiency of the antenna, along with an MS fit, are shown in Extended Data Figure~5. The error in the fits show that the maximally smooth models are better than 1\% accurate. These fits are used in the scaling of the antenna temperatures to derive sky spectra and this error is considered in the error analysis below.  The total efficiency of the antenna is a product of reflection efficiency $(1-|\Gamma|^2)$ and radiation efficiency $\eta_{\rm rad}$.  Using the MS fit models for total and reflection efficiencies we derive the radiation efficiency, which is also shown in the figure.

{\noindent \bf Data and its reduction.} SARAS~3 was deployed on lake sites where local transient RFI was found to be relatively infrequent and of low intensity.  
Another key consideration in site selection was the salinity and depth of water; these were measured to select sites where effects arising from reflections from the lake bed would not confuse CD/EoR detection.  SARAS~3 was deployed in Southern India on Dandiganahalli Lake (longitude $77.\!\!^\circ65981667$ East, latitude $+13.\!\!^\circ50896667$) during the nights of 2020 January 27--30, and on Sharavati backwaters (longitude $74.\!\!^\circ876027$ East, latitude $+13.\!\!^\circ992163$) during the nights of 2020 March 10--20.  Observing sessions were overnight from 10~pm to 7~am local time (Indian Standard Time).

16 complex spectra, with integration time 67.11~ms and spectral resolution 61~kHz, are acquired by the correlation spectrometer in each of the 6 instrument switch states as it cycles through. The real and imaginary spectra are Hampel filtered \cite{10.2307/2285666} to reject outliers beyond twice standard deviation.  Then the 16 spectra are averaged to yield six complex spectra, corresponding to the six switch states, for every 8.23~sec.  For the six spectra, an estimate of the measurement noise in each frequency channel is also recorded as metadata from the variance of the 16 measurements that are averaged. From every set of these six spectra, a double-differenced and bandpass-calibrated spectrum is computed and scaled to kelvin units.  Only the real component is useful, and represents a differential spectrum between the antenna temperature and that of the reference load.

Detecting and rejecting data that might be corrupted by RFI is done in multiple steps.  In the first step, individual 8.23-sec spectra are fitted with $10^{\rm th}$ order polynomials in the 40--90~MHz band and residuals are examined for outliers. A sufficiently high polynomial order is selected, considering the width of the band, so that plausible CD/EoR signal templates that might be present in the data would be fitted out and absent in the residual. Initial passes perform fits using L1 minimisation, so that the fit is less sensitive to large outliers, and later passes use L2 minimisation so that the fit follows the data in a least-squares sense \cite{RePEc:eee:ejores:v:73:y:1994:i:1:p:70-75}. Initial passes reject outliers exceeding $3\sigma$ of the noise, later passes reject deviations beyond $2 \sigma$.

In the second step, Legendre polynomials of order 15 are fitted to 8.23-sec spectra, and the residuals are progressively averaged over increasingly wide windows in time-frequency space, rejecting outliers exceeding $3\sigma$ at each selection of 2D smoothing window.  In time the maximum averaging is over 27.4~min; in frequency the maximum averaging is over 0.488~MHz. Inspection for outliers in 2D averaged data is aimed at detecting RFI that may persist in time exceeding the 8.23-sec basic cadence in calibrated spectra and/or be spread in frequency wider than the native resolution of 61~kHz.  Such RFI, if present, would be detectable with greater significance above the instrument noise when the data are averaged with a time-frequency window that matches the characteristic of the RFI; therefore, this search for RFI is done over a range of 2D smoothing in time-frequency space. 

With much of the obvious RFI rejected, the average power in the 40--90~MHz band and the RMS value of the noise in the spectra are examined versus time, to reject entire spectra that are outliers.  Odd numbered frequency channels are also rejected, since the use of the minimum 4-term window function on the time sequence before computing the Discrete Fourier Transform makes alternate channel data redundant \cite{2020JAI.....950006G}. 

An average spectrum is computed for each night. The RMS value of measurement noise in each frequency channel is also computed by propagating the noise through calibration and time averaging, accounting for the times rejected as RFI. Then a high $15^{\rm th}$-order polynomial fit is computed for each of the averaged spectra and these fits are subtracted out to yield residuals. Those spectral channels where this residual deviates by more than $2\sigma$ of that expected from measurement noise, and hence might be affected by RFI, are rejected. Frequency channels with substantial data rejection over the night, which might perhaps continue to have RFI at lower levels at times when it has gone undetected, 
are rejected for the entire night.

A source of self-generated internal RFI is the optical modulator's switched-mode power supply; this RFI is composed of a set of discrete, stable tones at a few frequencies. This unwanted component is almost completely cancelled in the correlation spectra by the differences; however, it appears in the auto-correlation spectra of the signals in the two receiver chains of the correlation receiver.  We Hampel filter the average auto-correlation spectrum with a sliding window, detect channels corrupted by this internal RFI, and reject the corresponding channel data in the cross-correlation spectra.

Next, we examine the average spectrum of each night for wide-band RFI.  The residuals obtained after fitting out a high-order polynomial from the average spectrum is examined for locations where at least five contiguous channels---each of 61~kHz width---all have positive deviations.  Such blocks of channels are rejected.

As a last step in the processing of the data, the physical temperatures of the reference, $T_{\rm ref}$, recorded over the observing duration are also added to the differential calibrated spectra to convert the measurements to absolute antenna temperature. These spectra are then corrected for the total efficiency of the antenna, to refer the spectra to sky domain.  

Admittedly, these spectra do contain some unwanted contributions from receiver noise and thermal emission from water beneath the antenna, scaled up by the total efficiency. The contribution from thermal emission of water, referred to the sky domain, is estimated to be $T_{\rm w}(1-\eta_{\rm rad})/\eta_{\rm rad}$, where $T_{\rm w}$ is the effective physical temperature of the water beneath the antenna and $\eta_{\rm rad}$ is the radiation efficiency of the antenna. Separately, receiver noise wave parameters are estimated from laboratory measurements, where the antenna was replaced with a 2-m low-loss coaxial cable and terminated with precision electrical short and open, and these are used to estimate the receiver noise spectrum $T_{\rm rec}$ in the measurement data. The derived contributions from water and receiver, referred to sky, are shown in Extended Data Figure~9.  It may be noted here that the receiver noise is negative because the phase change in its internal reflection at the antenna terminals results in partial noise cancellation and hence reduced noise figure for the receiver.  The estimated thermal emission from water and receiver noise are subtracted from the calibrated spectra referred to sky domain, to yield an estimate of the spectrum of the radio sky:  
\begin{equation}
    T_{\rm sky} = \frac{T_{\rm meas} + T_{\rm ref}}{\eta_{\rm rad} (1-|\Gamma|^2)} 
    - \frac{T_{\rm rec}}{\eta_{\rm rad} (1-|\Gamma|^2)}
    - \frac{T_{\rm w} (1-\eta_{\rm rad})}{\eta_{\rm rad}}.
\label{eq:caleqn}
\end{equation}
$T_{\rm meas}$ is the measurement data after bandpass calibration and scaling to kelvin units. 

At the observing sites, substantial RFI was encountered up to 55~MHz and the spectrum was also unusable above 85~MHz; consequently, the science band for the data from the observing sessions analysed herein is restricted to 55.5 to 84.5~MHz.

We reject data of an entire observing night if more than half of the data in that night is deemed by the above algorithms to be corrupted by RFI. Data in each of the remaining nights are divided into time blocks of 30~min each, and a single time-averaged spectrum is computed for each block.  We fit $6^{\rm th}$-order polynomials to these block-averaged spectra.  The residuals to the fits have a native resolution of 61~kHz and are smoothed, by convolution with Hann functions \cite{1978IEEEP..66...51H} with FWHM up to 2.8~MHz, and the RMS value of the residuals is recorded versus  FWHM.  For comparison, we construct two sets of mock spectra: both sets contain a model for the foreground radio sky plus measurement noise and the second set contains, in addition, the best-fitting profile that was found by Bowman et al. \cite{2018Natur.555...67B}. The foreground model is derived at each LST and for the observing site via a convolution of the SARAS~3 beam pattern and GSM to represent the spectrum of the radio sky in any celestial direction.  The measurement noise in each spectral channel is Gaussian random and of variance matched to that of the data in that block average. The two sets of mock spectra are processed the same way as the measurement data to provide a prediction for the RMS value of the residuals versus FWHM.  Each 30-min data block is qualified to be acceptable for science analysis only if the RMS value of its residual, smoothed to a resolution of 2.8~MHz FWHM, is within 2 standard deviations of the predictions.

Sky brightness temperature versus LST, as measured by the SARAS~3 instrument, is shown  in Extended Data Figure~6.  The qualified 30-min data blocks over the LST range of observing and from all the nights, with calibrations applied and receiver noise and thermal emission from water subtracted, are time averaged to form a single spectrum of the radio sky.

{\noindent \bf Model for the foreground and systematic calibration errors.} 
We constructed mock sky spectra using GSM and the SARAS~3 chromatic beam, and separately using GSM and a frequency-averaged SARAS~3 beam, which is achromatic. The mock spectra are in the science band of the SARAS~3 measurement, assume observations from the site where SARAS~3 was deployed and average over the LST range of the observations.  For the mock spectra that were made assuming an achromatic beam, the RMS value of the residual spectrum obtained from fitting a $3^{\rm rd}$ order polynomial is 0.7~mK; however, for the chromatic beam the RMS value of the residual is 6~mK for $3^{\rm rd}$ order and 0.4~mK for $4^{\rm th}$ order.  Therefore, to control the RMS value of systematics to be below 1~mK, a $4^{\rm th}$ order polynomial model is necessary for the foreground in the science band, given the small chromaticity of the SARAS~3 beam. 

The systematic errors in the total efficiency are estimated from the day-to-day variations in the efficiency estimates. The systematic errors arising from the use of GSM have been included by also computing this efficiency using GMOSS \cite{2017AJ....153...26S,2021MNRAS.502.4405B} as the sky model. The variation in the spectral profile of the receiver noise is computed from errors on fits for parameters of the receiver system \cite{2021ExANambissan}, which were determined while modeling the system response with the measurement equation.  We perturb the calibration terms in equation~(\ref{eq:caleqn}) to derive a large number of sky spectra $T^{\prime}_{\rm sky}$ that would reflect the scatter due to calibration errors.  Specifically, we compute instances of $T^{\prime}_{\rm sky}$:
\begin{equation}
    T^{\prime}_{\rm sky} = \frac{\eta (T_{\rm sky} + T_{\rm ref})}{\eta^{\prime}} 
    - \frac{(T_{\rm rec} - T^{\prime}_{\rm rec})}{\eta}
    - T_{\rm w} \left(\frac{(1-\eta_{\rm rad})}{\eta_{\rm rad}} - \frac{(1-\eta_{\rm rad}^{\prime})}{\eta_{\rm rad}^{\prime}}   \right),
\label{eq:2}
\end{equation}
\noindent where $\eta$ and $\eta_{\rm rad}$ are our best estimates for total and radiation efficiencies, which are estimated using all of the data, whereas $\eta^{\prime}$ and $\eta_{\rm rad}^{\prime}$ are drawn from sets representing the spread due to errors. Similarly, $T_{\rm rec}$ is our best estimate of receiver noise whereas $T^{\prime}_{\rm rec}$ is receiver noise drawn from the set representing the spread in this contribution due to errors in estimating the receiver system parameters.

In Extended Data Figure~7, we show mock spectra $T^{\prime}_{\rm sky}$ that are generated using one of GSM or GMOSS as the sky model, chromatic SARAS~3 beam, and allowing variations in the terms in equation~(\ref{eq:2}).  These spectra are representative of what might be obtained as measurements of the global sky as a result of systematic calibration errors.  Measurement noise has not been added. Also shown in the figure are residuals on fitting and subtracting polynomials of $3^{\rm rd}$, $4^{\rm th}$, $5^{\rm th}$, $6^{\rm th}$ and $7^{\rm th}$ order represented in log-temperature versus log-frequency.  The RMS values of the residuals represent the systematic error that might be expected if polynomials of the corresponding order were selected to model the foreground.  In order to control the RMS value of systematic calibration errors to be subdominant to the profile found by Bowman et al.\cite{2018Natur.555...67B}, the foregrounds plus calibration systematics need modelling with at least $6^{\rm th}$ order polynomials.

We added the profile found by Bowman et al.\cite{2018Natur.555...67B} to a GSM model of the global sky in the science band of the SARAS~3 observations and subtracted polynomials of different orders to get residuals.  These are also shown in the figure and  represent what might be expected as residuals of SARAS~3 measurements if the profile found by Bowman et al.\cite{2018Natur.555...67B} were present in the data.  In spectra with native resolution of 61~kHz, the SARAS~3 data has instrument noise with RMS value of 213~mK.  Adding variance from systematics and measurement noise, we find that SARAS~3 data has substantially greater sensitivity to detect the profile found by Bowman et al.\cite{2018Natur.555...67B} if the foreground plus calibration systematics were modelled using a $6^{\rm th}$ order polynomial instead of $7^{\rm th}$ or $5^{\rm th}$.

For this adopted $6^{\rm th}$ order polynomial model, we have derived the magnitude of systematic errors by assuming errors in estimations of just one causative factor at a time.  We thus separately examine errors from total efficiency, thermal emission from water and receiver noise. Extended Data Figure~9 shows these individual contributions. The RMS value of the residuals resulting from errors in each contributor are found to be below 1~mK.

We estimate the robustness of inference about the profile based on corrections in equation~\ref{eq:2}. This is evaluated by computing scale factor associated with the best-fit profile (see equation~\ref{eq:model}), by forming the spectrum using different corrections as shown in Extended Data Figure~9. We also form spectra corresponding to extreme scenarios, where we omit one or all of these corrections. We find that the uncertainty in these corrections result in a scatter of $\pm 0.03$ in the recovered scale factor.  Additionally, to account for systematics that might not be captured by the above modelling, we compare the residuals of fits to $6^{\rm th}$ order polynomial to that from higher orders. We fit $6^{\rm th}$ and higher order polynomials to the SARAS~3 spectrum and compute the RMS values of the residuals, after smoothing to 1.4~MHz frequency resolution; these are depicted in panel~(d) of Extended Data Figure~8.  The RMS values have been corrected for the reduction in degrees of freedom corresponding to the order of the fitting polynomial. The smoothing scale is chosen so that any ripples arising due to reflections in the optical fibers are cancelled.  The RMS values are consistent suggesting that there is no evidence for systematic errors in these residuals, including the residual of the $6^{\rm th}$ order fit. Nevertheless, we use the residuals of the higher-order fit to compute 95\% confidence lower limit of 39.7~mK on RMS value of measurement noise in the residuals of $6^{\rm th}$ order fit, and hence derive an upper limit of 8.6~mK on any systematic errors in the residuals of $6^{\rm th}$ order fit.  The corresponding uncertainty on scale factor is found to be $\pm 0.32$. As discussed in the main section, these potential systematic errors have been included in the error budget and the significance of results derived have been reduced to reflect such errors.

We also performed Bayesian model selection on the calibrated SARAS~3 spectrum of the global sky, to ascertain the polynomial order favoured by the data for describing the foregrounds plus any calibration systematics.  The model selection allows for the profile found by Bowman et al. \cite{2018Natur.555...67B}, if present in the data, to be separately modelled and is thus unbiased. We compute Bayesian evidence \cite{2015MNRAS.453.4384H} and Bayesian information criterion (BIC) \cite{1978AnSta...6..461S} as the metrics for selecting the order of the polynomial model. The model with higher Bayesian evidence or lower BIC is preferred. In order to assess the significance of these metrics, we use Jeffery's scale \cite{10.2307/2291091}.  We compute Bayesian evidence and BIC to compare between two types of modelling of the SARAS~3 data: (a) a joint model, where the spectrum is fit to a polynomial of order $N$ in log-temperature versus log-frequency, plus the profile found by Bowman et al. with a free scaling parameter, and (b) modelling where the spectrum is fit to a plain polynomial of order $N$.

In Extended Data Figure~8 we compare the metrics for these two types of models and for a wide range of polynomial orders.  Bayesian evidence and BIC strongly require that the modelling be done using at least a $6^{\rm th}$-order polynomial.  Within errors, Bayesian evidence has no preference between models that include polynomials of orders between six and eight; however, BIC clearly selects models with polynomial of order six.  We provide the evidence values, along with errors, and BIC in Extended Data Table~1.

Bayesian evidence and BIC consistently prefer a plain polynomial model over that which includes the profile found by Bowman et al. \cite{2018Natur.555...67B}, with a free parameter. This is consistent with our finding in the Main section of the manuscript that a joint fit that includes the profile found by Bowman et al. \cite{2018Natur.555...67B} results in a scale factor of 0. The preference of both metrics---Bayesian evidence and BIC---for a $6^{\rm th}$-order polynomial-only model over a joint model of any order is additional evidence for rejecting the profile found by Bowman et al. \cite{2018Natur.555...67B}.

We have also considered the use of MS polynomial functions to represent the data.  For this model, the residual has RMS value about 1~K and the metric log(BIC) is 12.4; therefore, the MS polynomial model for foreground is rejected.

For systematic calibration errors in residuals to be subdominant, and for optimum detection of the profile found by Bowman et al.\cite{2018Natur.555...67B}, if present, SARAS~3 spectra of the global sky require modeling of foreground plus systematic errors using $6^{\rm th}$ order polynomials.  

{\noindent \bf Rejection confidence in the parameter space of the profile found by Bowman et al.\cite{2018Natur.555...67B}}
The temperature spectrum measured using the EDGES instrument was found to have structure that was modelled as a flattened Gaussian profile \cite{2018Natur.555...67B}.  Markov Chain Monte Carlo (MCMC) \cite{2013PASP..125..306F} analysis of that spectrum yielded confidence intervals on the parameters. The result of that MCMC analysis provides a sampling of the parameter space with density that encodes the likelihood, and in Extended Data Figure~10, panel (a), we show the corresponding profiles. Their best-fitting profile, which was rejected with 95.3\% confidence in the analysis in the main text, is also shown. Here we discuss the confidence in detail considering the distribution in the parameters. 

We have used this distribution of profile parameters and generated a substantially larger number of MCMC walkers by allowing for an additional scale factor parameter for each of their profiles.  We then evolved the walkers for each profile, using a $6^{\rm th}$ order polynomial to represent the foreground and calibration systematics in this sky spectrum, to model the SARAS~3 spectrum.  The analysis for each walker thus followed the same method described in the main text for the best-fitting profile of Bowman et al.\cite{2018Natur.555...67B}. The scale factor distributions from our suite of walkers, considering all of the walkers for all of the profiles together, are naturally weighted by the likelihood distribution defined by the data from the EDGES instrument, and constrained by the data from SARAS~3. The scale factors from all walkers, combined to form a single distribution, has the distribution shown in panel (b). The mean scale factor is 0.07 and the profile set found by Bowman et al.\cite{2018Natur.555...67B} as a whole is rejected with 90.4\% confidence. In panel (c) of the figure we display the distribution of most-likely scale factors corresponding to each of the profiles in panel (a), showing that for none of the profiles is a scale factor consistent with unity, and is therefore disfavored by the SARAS~3 data.  In Extended Data Figure~11 we show the distribution of rejection confidence in the profile parameter space. 

{\noindent \bf Jackknife tests. } 
The data from the SARAS~3 instrument were divided and 
jackknife tests performed to examine the dependence of MCMC analysis results on statistical and systematic errors. With these tests, if we are not limited by the errors, we expect the best-fit scale factor $s$ to vary with noise realisation, while being consistent with zero in all cases. First, the spectral data with 8.23-sec integration time were divided into even and odd time samples, and averaged separately to form two independent measurements of the sky spectrum.  MCMC analysis that modelled the spectrum using a $6^{\rm th}$ order polynomial plus a scale factor $s$ times the best-fit profile found by Bowman et al.\cite{2018Natur.555...67B} gave best-fit scale factor to be $-0.63 \pm 0.81$ and $0.66 \pm 0.75$ for the two data sets; both are consistent with $s=0$ within $1 \sigma$ uncertainty. Second, the selected data set was sliced into two different sets of nights, corresponding to early and later parts of the observing schedule. MCMC analysis of these data gave best-fit scale factor $-0.17 \pm 0.72$ and $0.23 \pm 0.70$, which are again both consistent with $s = 0$ within their $1 \sigma$ errors. Naturally, these data with reduced integration times yield estimates for $s$ with larger errors.  

\bgroup
\def\arraystretch{2}
\setlength{\tabcolsep}{25pt}
\begin{table}[h]
\centering
\begin{tabular}{|c c c c|}
 \hline
 \textbf{S.N.} & \textbf{Model} & \textbf{log(evidence)} & \textbf{BIC}\\ [0.5ex] 
 \hline
 1 & Joint(4) & $ -789.09 \pm 0.88$ & -1290.30 \\ 
 \hline
 2 & Poly(4) & $ -794.50 \pm 0.61$ & -1283.88 \\
 \hline
 3 & Joint(5) & $-785.23 \pm 0.53$ & -1295.89 \\
 \hline
 4 & Poly(5) & $-782.10 \pm 0.56$ & -1301.65 \\
 \hline
 5 & Joint(6) & $-724.02 \pm 0.52$ & -1405.90 \\ 
 \hline
 6 & Poly(6) & $-719.29 \pm 0.53$ & -1412.06 \\
 \hline
 7 & Joint(7) & $-725.62 \pm 0.50$ & -1403.84 \\
 \hline
 8 & Poly(7) & $ -721.25 \pm 0.51$ &  -1406.86\\
 \hline
 9 & Joint(8) & $-723.29 \pm 0.47$ & -1402.26 \\ 
 \hline
 10 & Poly(8) & $-719.03 \pm 0.47$ & -1403.36\\
 \hline
 11 & Joint(9) & $-727.86 \pm 0.47$ & -1396.28 \\
 \hline
 12 & Poly(9) & $-725.13 \pm 0.48$ & -1399.83 \\
 \hline
 13 & Joint(10) & $-727.79 \pm 0.45$ & -1390.12 \\ 
 \hline
 14 & Poly(10) & $-727.38 \pm 0.46$ & -1393.84 \\
 \hline
 15 & Joint(11) & $-740.27 \pm 0.47$ & -1385.70 \\ 
 \hline
 16 & Poly(11) & $-733.24 \pm 0.46$ & -1388.08 \\ 
 \hline
 17 & Joint(12) & $-742.29 \pm 0.46$ & -1379.81 \\ 
 \hline
 18 & Poly(12) & $-742.00 \pm 0.47$ & -1383.39 \\ [1ex] 
 \hline
\end{tabular}
\caption{Values of log$_{10}$(Bayesian evidence) and Bayesian Information Criterion (BIC) for joint models, where the model is a polynomial plus the profile found by Bowman et al. with a free scaling factor, and plain polynomial models.  The models are labelled as either `Joint' or `Poly' for the two cases, respectively, with the order of the polynomial given within brackets.
}.
\label{tab:evidence}
\end{table}
\egroup

\vspace{1cm}

\noindent {\bf Data availability statement.} The processed spectrum and code that supports the findings of this study are available from the corresponding authors on reasonable request.

\vspace{1cm}

\noindent {\bf Code Availability statement.} The dataset is stored and pre-processed in miriad (\url{https://www.atnf.csiro.au/computing/software/miriad/}). Later stages of analysis are done in python, extensively using publicly available routines in scipy (\url{https://scipy.org/}), numpy (\url{https://numpy.org/}) and matplotlib (\url{https://matplotlib.org/}). MCMC analysis was done using \texttt{emcee} package (\url{https://github.com/dfm/emcee}), and Bayesian evidence was computed using \texttt{PolyChord} (\url{https://github.com/PolyChord/PolyChordLite}). Other processing and analysis steps were performed using custom code, which is available from the corresponding author upon reasonable request.

\clearpage

%% file: extended_data.tex
\begin{figure}
\centering
\includegraphics[width=1.0\linewidth]{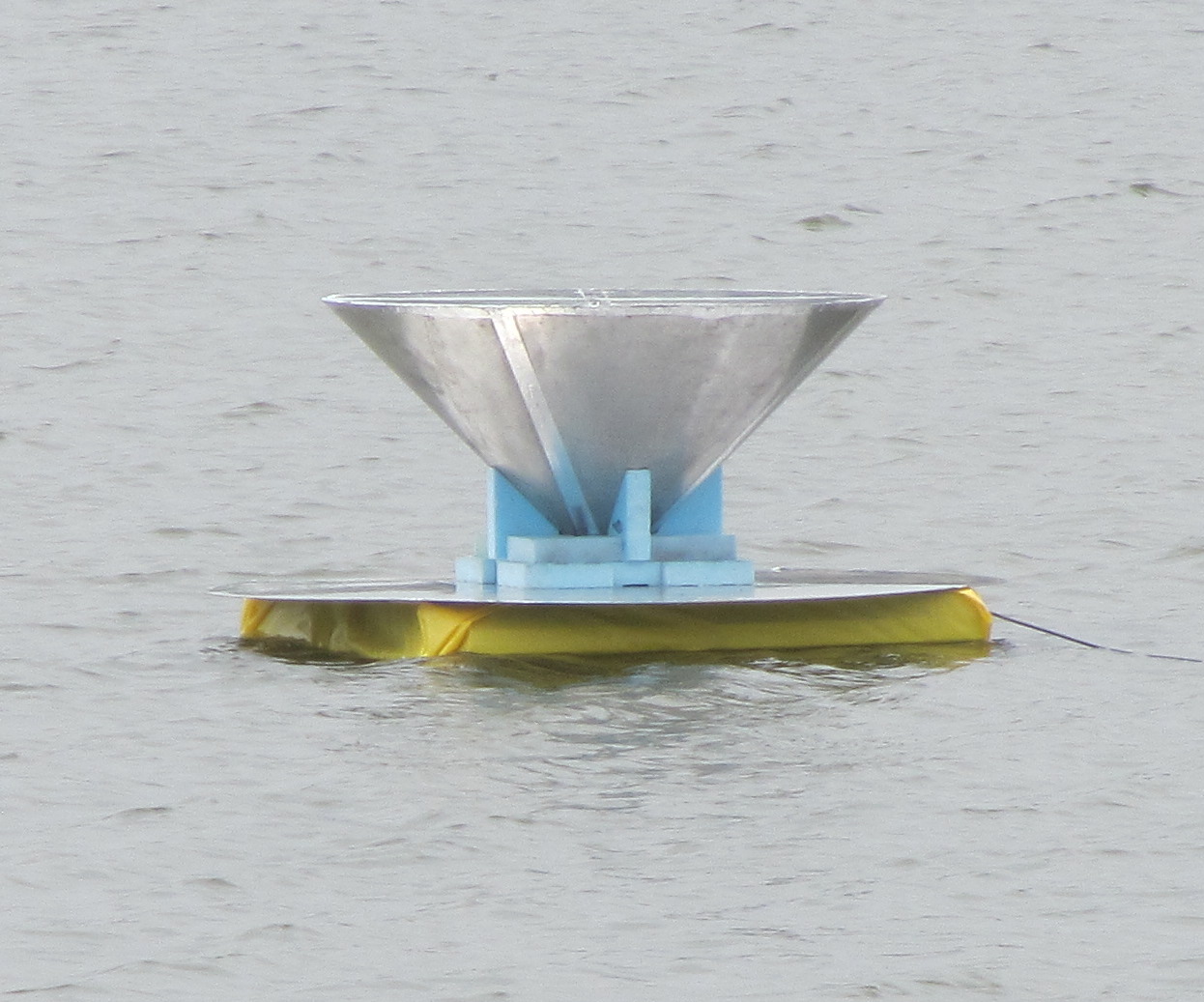}
\caption*{{\bf Extended Data Figure 1: The SARAS~3 antenna.} The monocone antenna is shown, floating on water on its raft.  
The antenna electronics is in an enclosure beneath the antenna ground plane and within the raft; power is derived locally from Li-ion battery packs within the enclosure.  Multi-core fibre optic cables connect the antenna to the analogue signal conditioning unit (ASCU) in the base station on shore.
}
\label{fig:antenna}
\end{figure}

\clearpage

\begin{figure}
\centering
\includegraphics[width=1.0\linewidth]{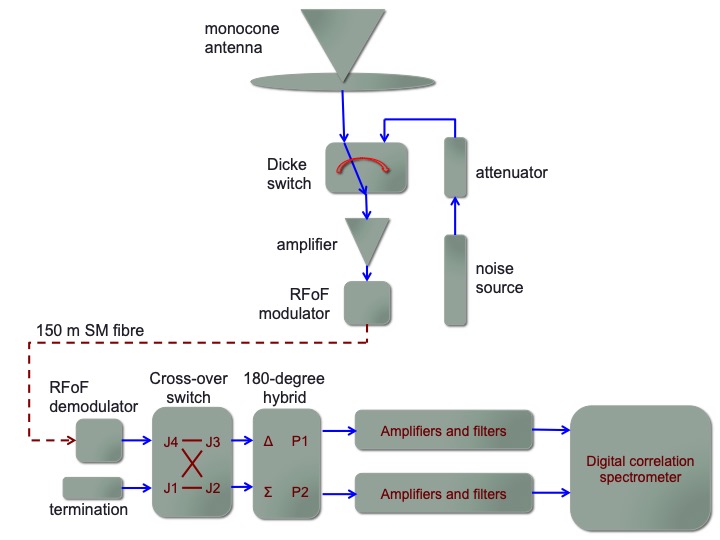}
\caption*{{\bf Extended Data Figure 2: The architecture of the SARAS~3 instrument.} The building blocks and signal flow within the instrument is shown from the antenna to the digital spectrometer.}
\label{fig:instr_arch}
\end{figure}

\clearpage

\begin{figure}
\centering
\includegraphics[width=0.85\linewidth]{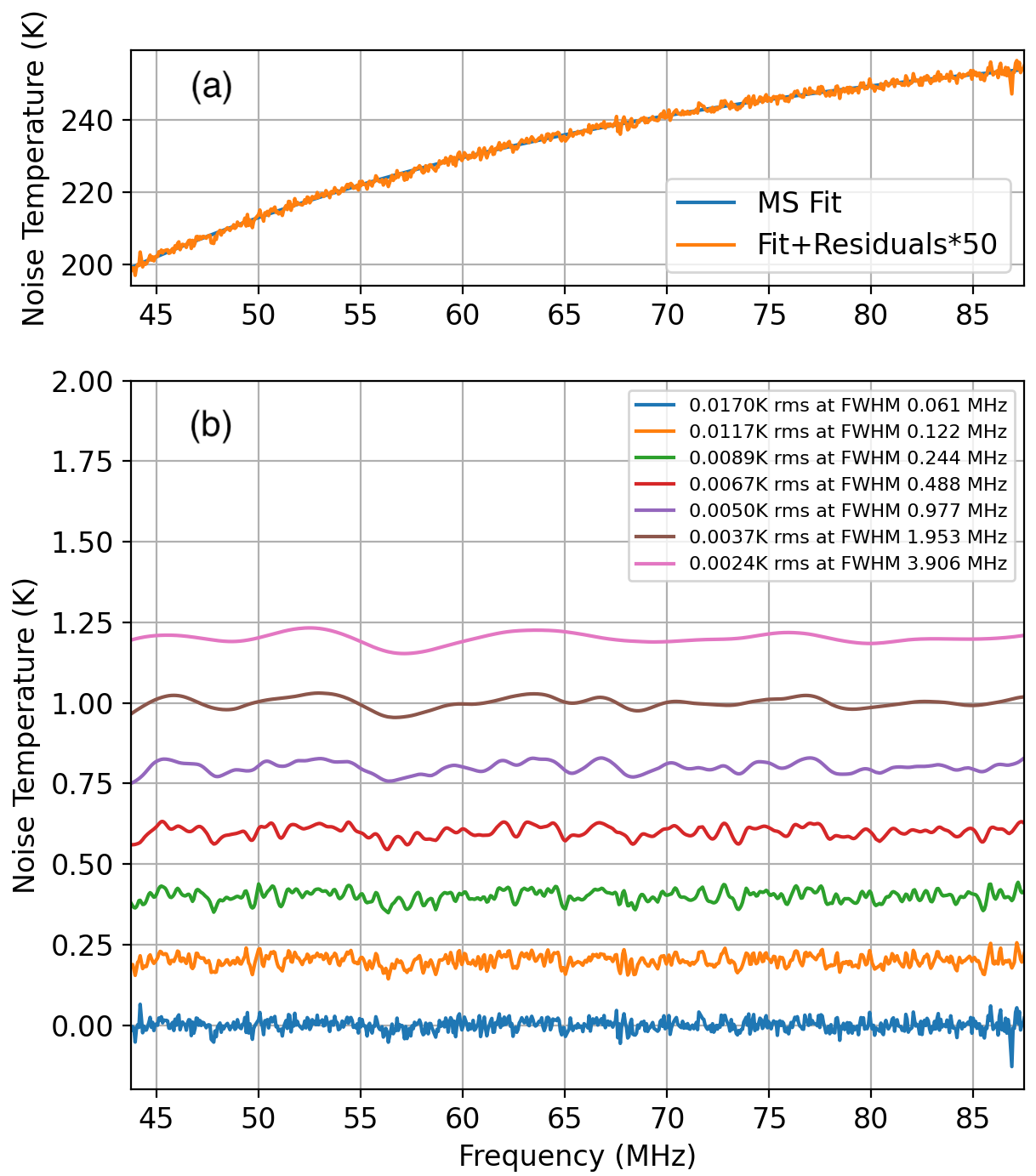}
\caption*{{\bf Extended Data Figure 3: Laboratory verification of the receiver. }  Panel (a) shows the calibrated spectrum measured in the laboratory, in a long duration test, with the antenna replaced by an antenna simulator that approximates its reflection coefficient.   A maximally smooth function fit to the spectrum is shown using a red dashed line; the data, with deviations from the fit magnified by factor 50, is shown as a blue continuous line. Panel (b) shows the residuals smoothed to successively lower resolutions, using a Hann function, from the native 61~kHz to 3.906~MHz.  The FWHM of the smoothing functions and the RMS values of the residuals on smoothing is in the legend. For clarity, the traces have been offset vertically and magnified by the ratio of RMS value after smoothing to the RMS value at native resolution.}
\label{fig:asim_fit}
\end{figure}

\clearpage

\begin{figure}
\centering
\includegraphics[width=1.0\linewidth]{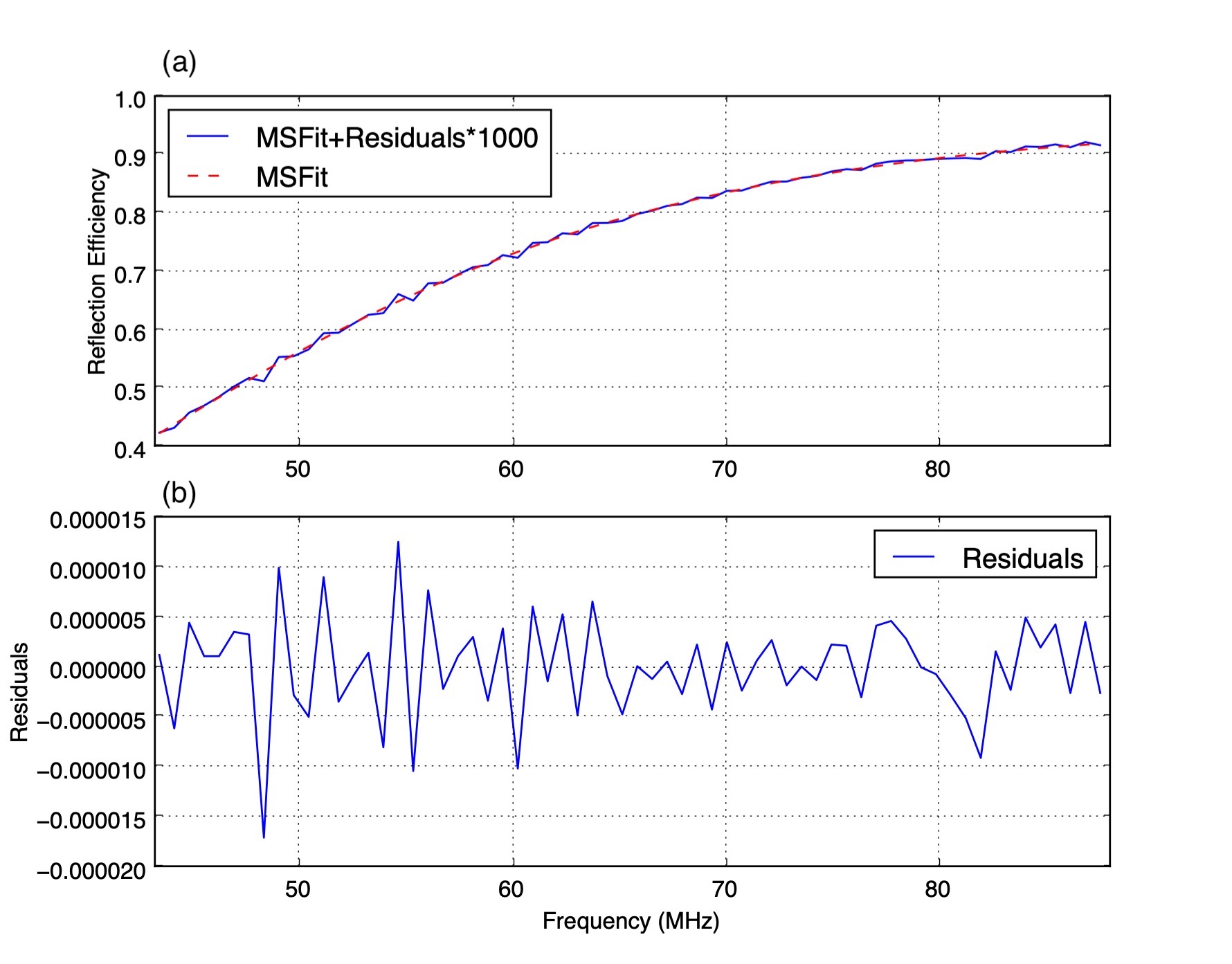}
\caption*{{\bf Extended Data Figure 4: Antenna reflection efficiency.} Panel (a) shows the fit of a maximally smooth function to the reflection efficiency, as a red dashed line. Overlaid is the measurement, as a continuous blue line, with the difference between the fit and measurement magnified by factor 1000 for clarity.  Panel (b) shows the difference between the measurement and MS fit.}
\label{fig:refeff_2}
\end{figure}

\clearpage

\begin{figure}
\centering
\includegraphics[width=0.7\linewidth]{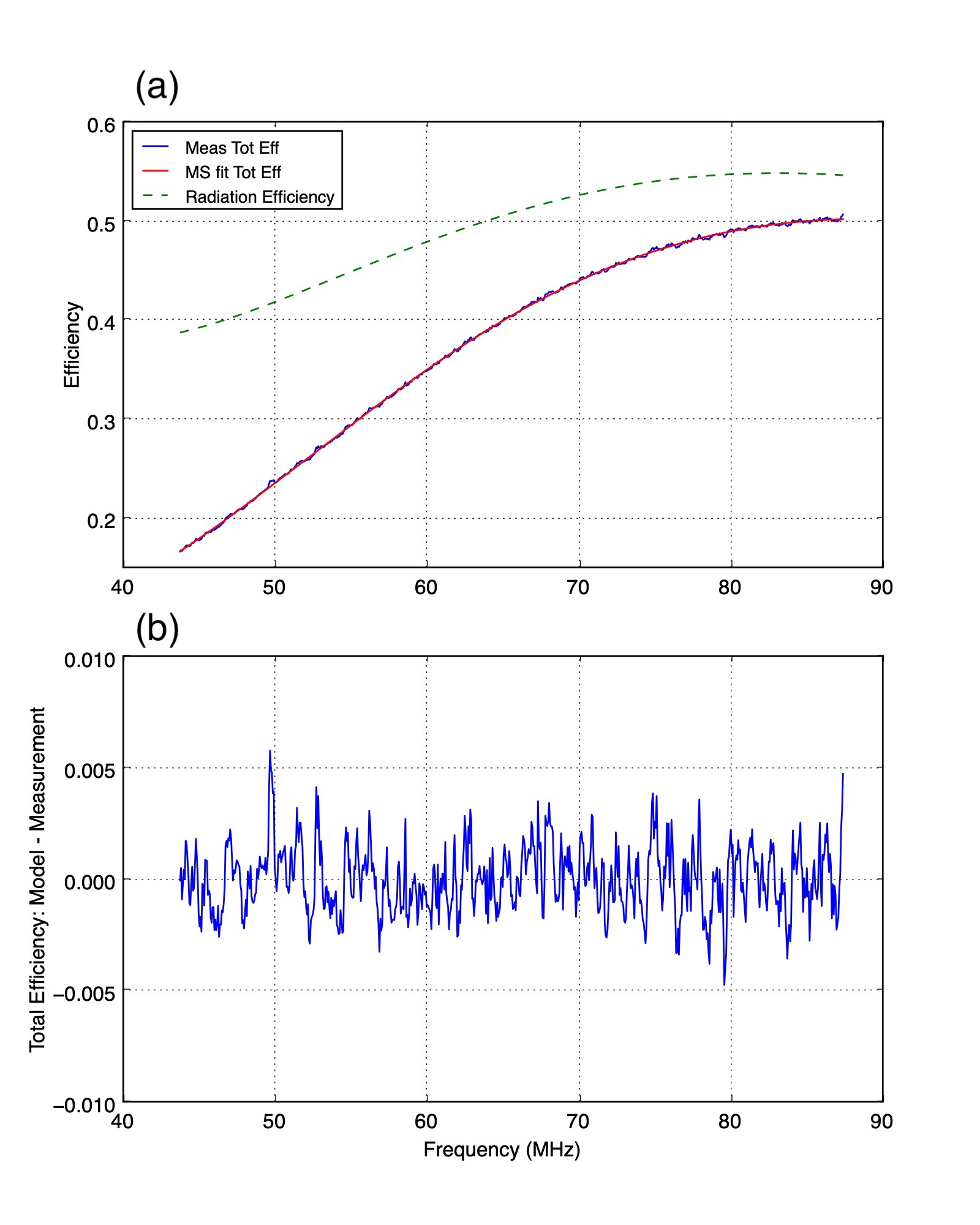}

\caption*{{\bf Extended Data Figure 5: Antenna total and radiation efficiencies.} Panel (a) shows the total and radiation efficiencies of the instrument as measured on water in the Sharavati backwaters, where most of the observations were made. Maximally smooth function fit to the total efficiency is overlaid on the measurement and the difference between the best-fit MS function and measurement is shown in panel (b).}
\label{fig:TotEff}
\end{figure}

\clearpage

\begin{figure}
\centering
\includegraphics[width=0.9\linewidth]{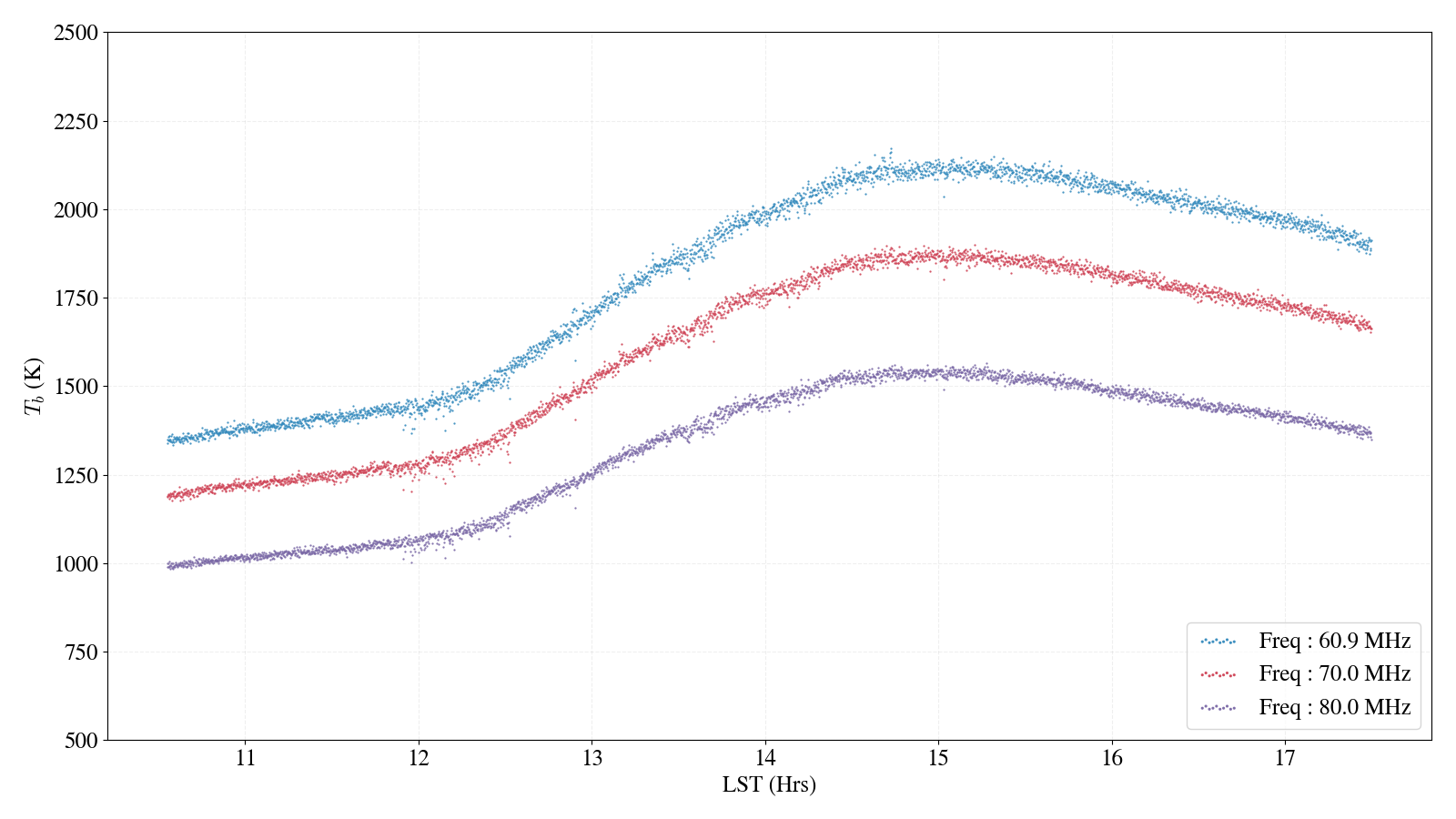}
\caption*{{\bf Extended Data Figure 6: Antenna temperature versus LST. } Measured sky brightness temperature versus local sidereal time (LST), at a set of frequencies across the science band.  The brightness temperature at any time is the global radio sky viewed by the antenna beam pattern of the instrument. }
\label{fig:LST_vs_pow}
\end{figure}

\clearpage

\begin{figure}
\begin{center}
\includegraphics[scale=0.48]{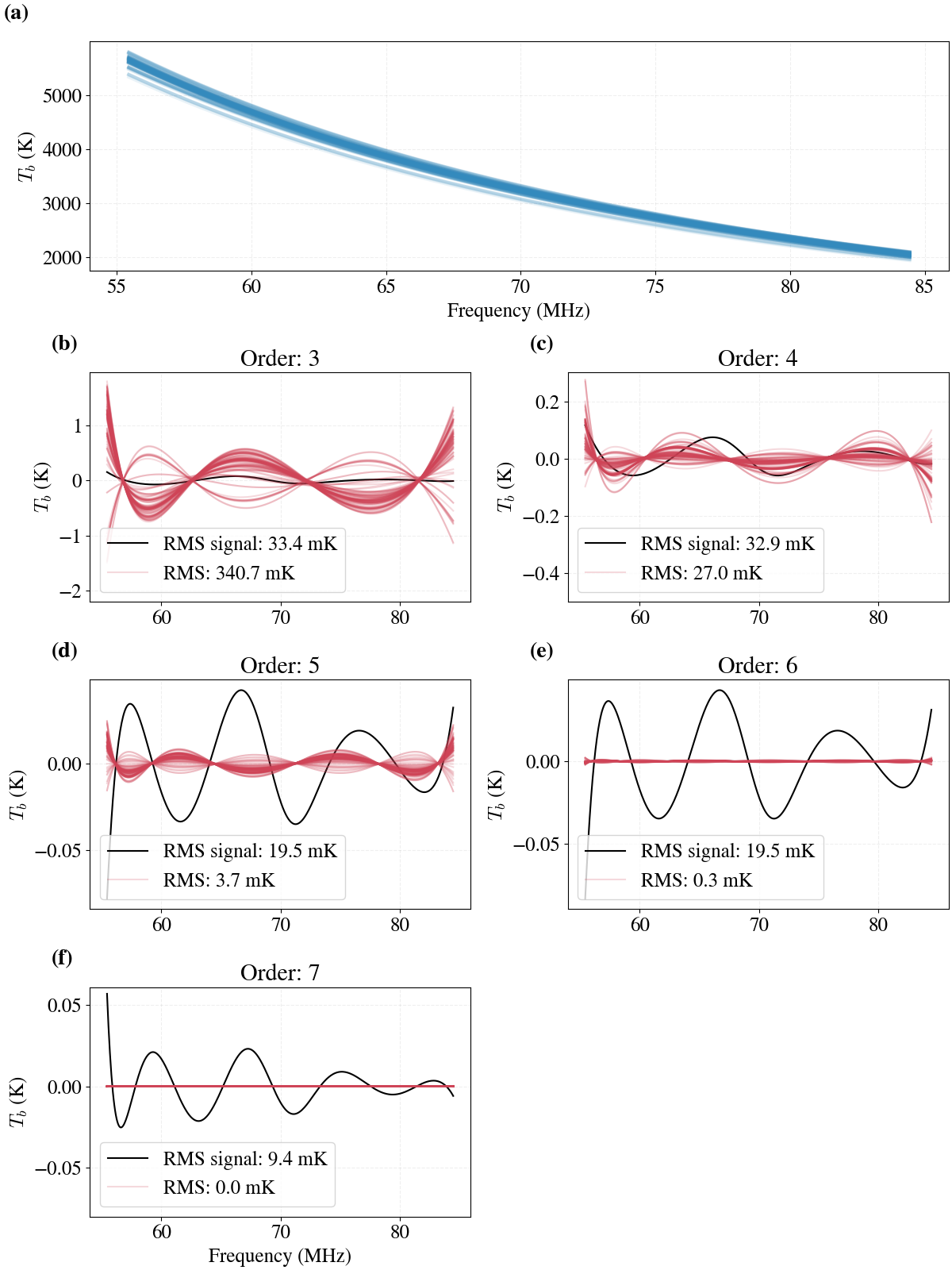}
\caption*{{\bf Extended Data Figure 7: Systematic calibration errors. }  Panel (a) shows mock sky spectra constructed using erroneous calibrations for efficiencies, water thermal emission and receiver noise, with errors reflecting the uncertainties in their derivations. Panels (b), (c), (d), (e) and (f) show, in red, the systematic residuals that might result due to calibration errors and using  $3^{\rm rd}$, $4^{\rm th}$, $5^{\rm th}$, $6^{\rm th}$ or $7^{\rm th}$ order polynomials to represent the foreground. For comparison we also show, in black, the residual expected if the global sky spectrum contained the profile found by Bowman et al.\cite{2018Natur.555...67B}.}
\label{fig:add_log_log}    
\end{center}   
\end{figure}

\clearpage

\begin{figure}
\begin{center}
\includegraphics[scale=0.44]{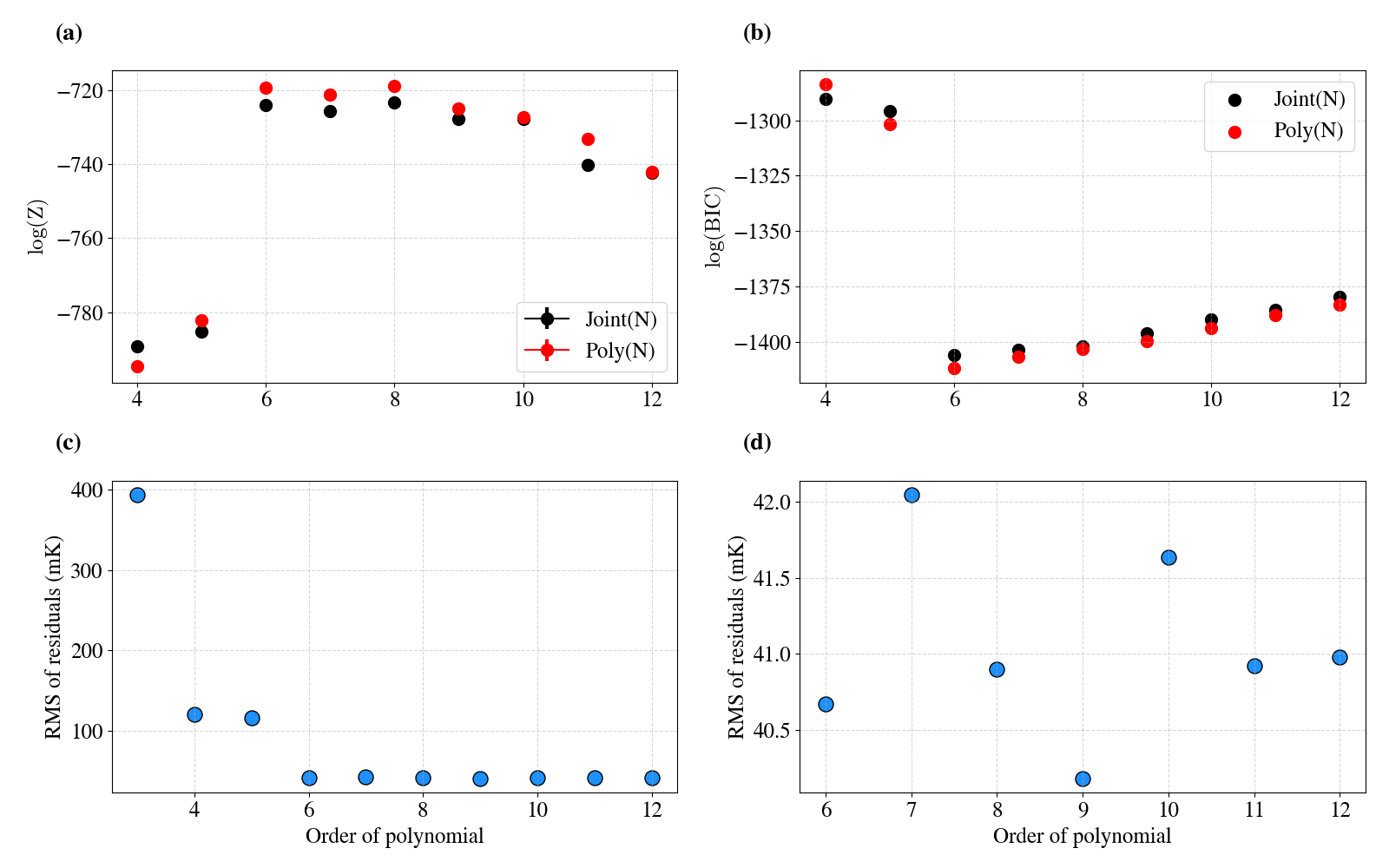}
\caption*{{\bf Extended Data Figure 8: Model selection.}
    {
    Panel (a) shows the logarithm of Bayesian evidence versus the order of the polynomial: in black symbols is shown the case where the sky spectrum measured by SARAS~3 is modelled as a polynomial plus the profile found by Bowman et al. with a free scaling factor, and in red symbols is shown the case where the model is a plain polynomial.  Higher the evidence, more preferred is the model. Panel (b) shows Bayesian Information Criterion: here models with lower values are preferred. Panel (c) shows the RMS value of the residuals smoothed to 1.4~MHz resolution: the spectrum is consistent with measurement noise for $6^{\rm th}$ and higher orders; panel (d) shows exclusively the RMS values for order 6 and above.
    }}
\label{fig:model_selection}    
\end{center}   
\end{figure}

\clearpage

\begin{figure}
\begin{center}
\includegraphics[scale=0.35]{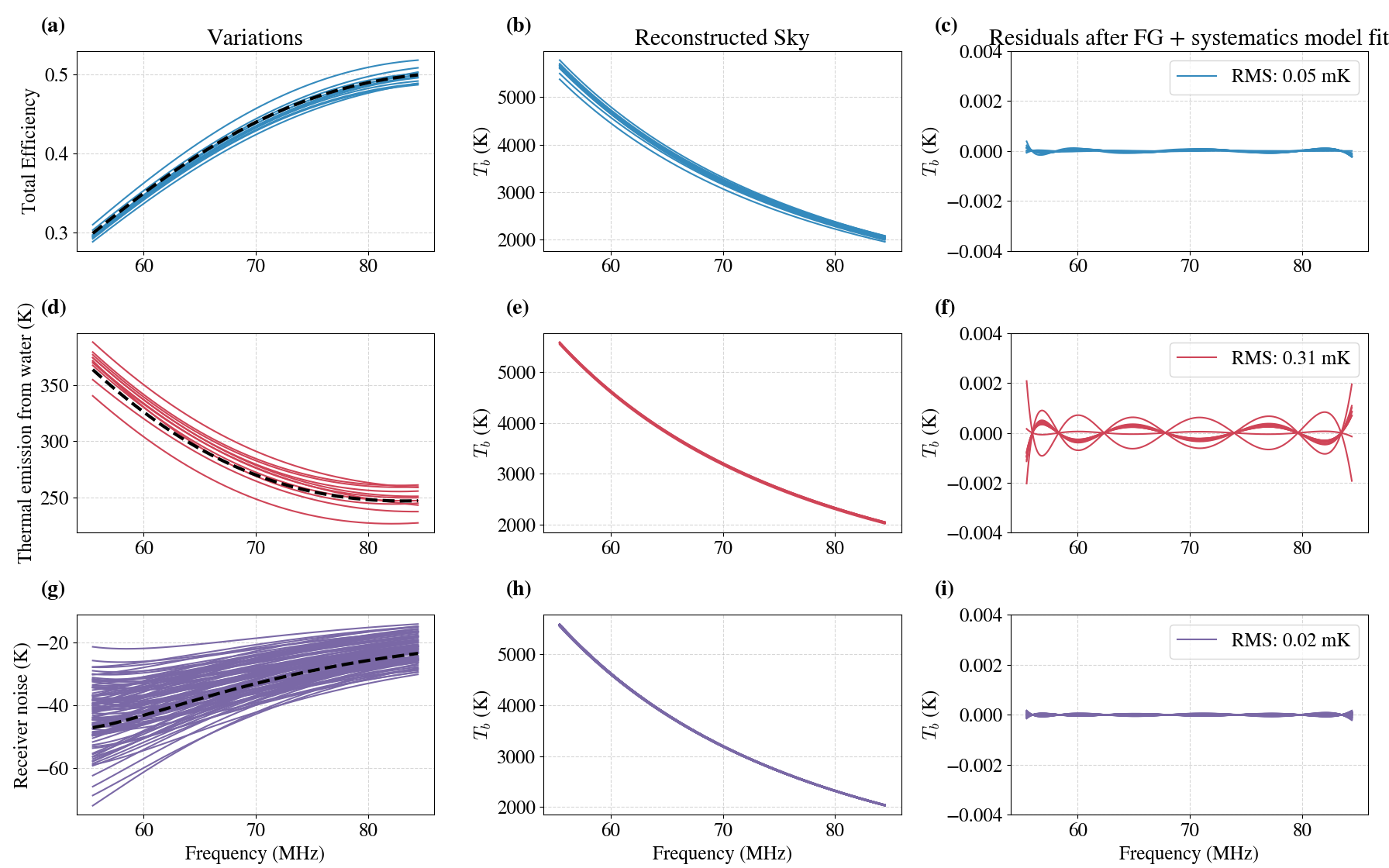}
\caption*{{\bf Extended Data Figure 9: Breakdown of systematic calibration errors.}  Mock spectra of the global sky are generated using GSM and GMOSS, using the SARAS~3 beam, and processed to impress systematic errors in calibrations for total efficiency (panels a, b \& c), water thermal emission (panels d, e \& f) and receiver noise (panels g, h \& i).  Calibrations with errors that span the expected distributions are shown in the first column; our best-estimate calibrations are also shown using black dashed lines.  Mock spectra with these calibration errors are shown in the second column.  Residuals from fitting and subtracting out $6^{\rm th}$-order polynomials from the mock spectra are shown in the last column.} 
\label{fig:6_order_errors}    
\end{center}   
\end{figure}

\clearpage

\begin{figure}
\begin{center}
\includegraphics[scale=0.45]{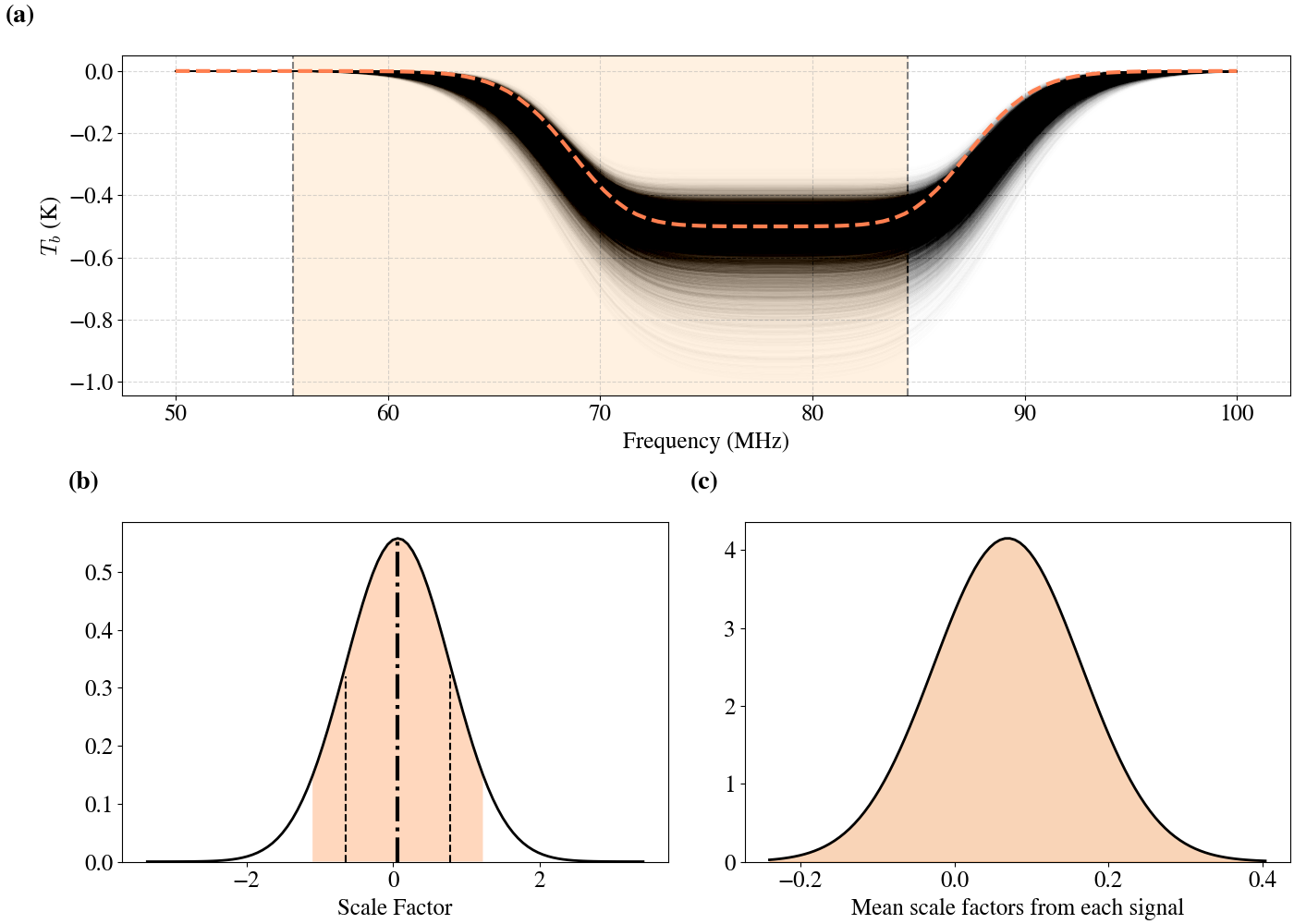}
\caption*{{\bf Extended Data Figure 10: Deriving confidence of rejection for the profile found by Bowman et al.\cite{2018Natur.555...67B}} Panel (a) shows flattened Gaussian profiles with parameters in the ranges charted by Bowman et al.\cite{2018Natur.555...67B}; the relative density in profile space displays their relative likelihood.  The dashed orange line shows the best-fitting profile; the vertical lines and shaded domain demarcate the band used in our analysis. Lower panels show the result of MCMC analysis using the SARAS~3 data that determines the distribution of scale factors associated with each of these profiles: (b) shows distribution of scale factors for the whole class and (c) shows the distribution of the mean scale factor corresponding to each of the profiles. The shaded region in panel (b) spans $5^{\rm th}-95^{\rm th}$ percentile values, and dash-dotted black line represents the mean. The dashed black lines represent $1\sigma$ confidence intervals. The profile set in panel (a) as a whole is rejected with 90.4\% confidence.} 
\label{fig:atlas}    
\end{center}   
\end{figure}

\clearpage

\begin{figure}
\begin{center}
\includegraphics[scale=0.35]{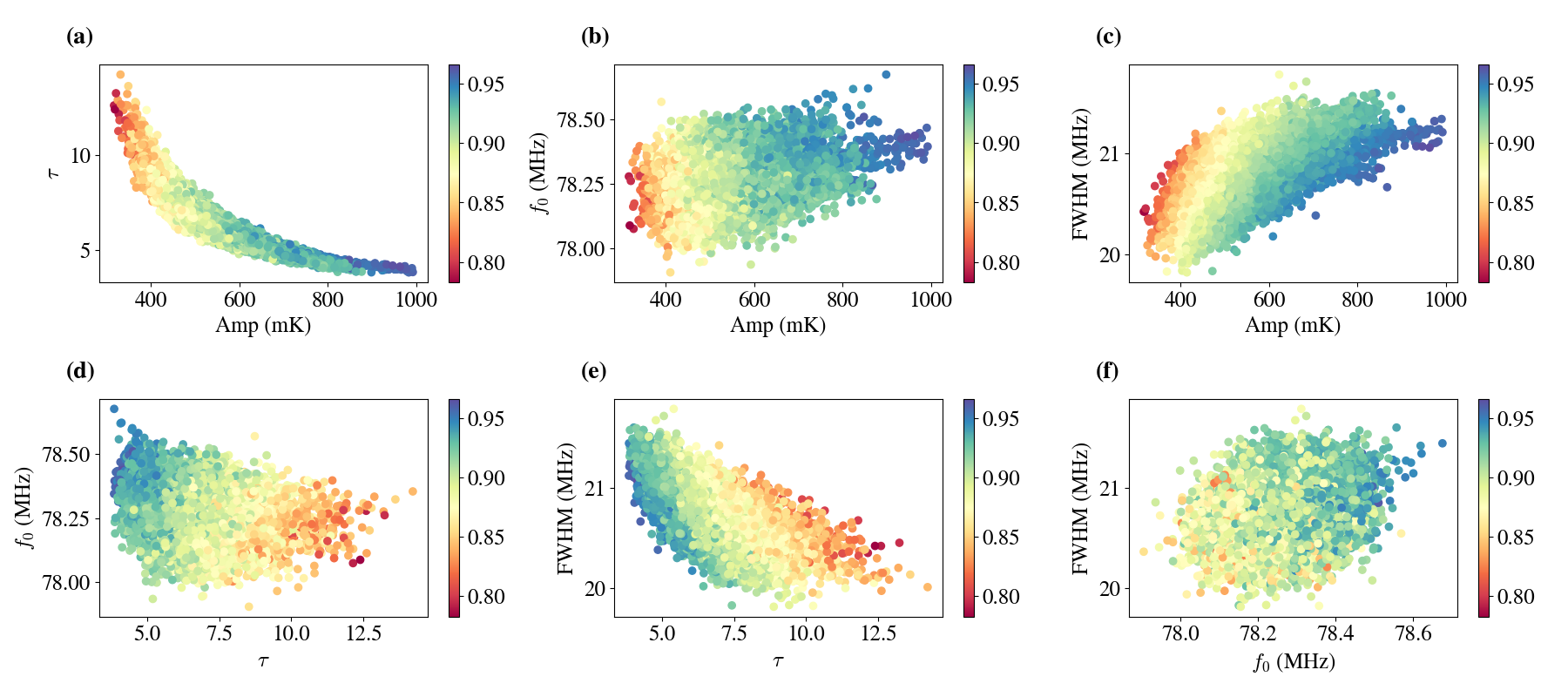}
\caption*{{\bf Extended Data Figure 11: Rejection confidence in the space of profile parameters. }  Scatter plots corresponding to the different parameters are shown in separate panels, colour coded to show the significance with which the different parameter spaces are rejected by our MCMC analysis using the SARAS~3 data. The parameters describing the signal are the same as in Bowman et al.: amplitude (in mK), central frequency $f_0$ (in MHz), full width at half maximum (FWHM) of the profile (in MHz) and the flattening parameter $\tau$.} 
\label{fig:atlas2}    
\end{center}   
\end{figure}